\documentclass[pra,aps,twocolumn,superscriptaddress,nofootinbib,nopacs]{revtex4} 

\usepackage{graphicx}
\usepackage{psfrag}
\usepackage{epsfig}
\usepackage{pst-all}
\usepackage{gensymb}
\usepackage{bm}
\usepackage{bbm}
\usepackage{braket}
\usepackage{dcolumn}
\usepackage{amsmath}
\usepackage{amssymb}
\usepackage{amsfonts}
\usepackage{epsfig}
\usepackage{mathrsfs}
\usepackage{mathtools}
\usepackage{lmodern}
\usepackage{indentfirst}
\usepackage{color}
\usepackage[T1]{fontenc}
\usepackage{comment}
\usepackage{xcolor}
\usepackage[colorlinks,linkcolor=blue,citecolor=blue,urlcolor=blue,hyperindex,driverfallback=dvipdfm]{hyperref}
\usepackage{soul}
\usepackage{float} 
\usepackage{enumerate}
\usepackage[overload]{empheq}
\newcommand*{\widebox}[2][0.5em]{\fbox{\hspace{#1}$\displaystyle #2$\hspace{#1}}}

\def\ii{{\rm i}}  \def\ee{{\rm e}}
\def\Ree{{\rm Re}}  \def\Imm{{\rm Im}}
\def\rb{{\bf r}}  \def\Rb{{\bf R}}    \def\vb{{\bf v}}
\def\xx{\hat{\bf x}}  \def\yy{\hat{\bf y}}  \def\zz{\hat{\bf z}}    \def\rr{\hat{\bf r}}
  
\def\RR{\hat{\bf R}}  
\def\kb{{\bf k}}    
  \def\Gb{{\bf G}}  \def\qb{{\bf q}}  \def\gb{{\bf g}}    
  \def\fb{{\bf f}}  \def\Sb{{\bf S}}    \def\qb{{\bf q}}  
\def\me{m_{\rm e}}  
\def\Eb{{\bf E}}  \def\Bb{{\bf B}}    \def\Ab{{\bf A}}
\def\pb{{\bf p}}  \def\jb{{\bf j}}  
  
\def\blue{\color{blue}}      
\def\uu{\hat{\bf u}}  \def\w{\omega}

\begin{document} 

\def\bibsection{\section*{\refname}} 

\title{Modulation of Cathodoluminescence Emission by Interference with External Light
}


\author{Valerio~Di~Giulio}
\affiliation{ICFO-Institut de Ciencies Fotoniques, The Barcelona Institute of Science and Technology, 08860 Castelldefels (Barcelona), Spain}
\author{Ofer~Kfir}
\affiliation{IV Physical Institute, Solids and Nanostructures, University of G\"ottingen, 37077 G\"ottingen, Germany}
\affiliation{Max Planck Institute for Biophysical Chemistry (MPIBPC), 37077 G\"ottingen, Germany}
\author{Claus~Ropers}
\affiliation{IV Physical Institute, Solids and Nanostructures, University of G\"ottingen, 37077 G\"ottingen, Germany}
\affiliation{Max Planck Institute for Biophysical Chemistry (MPIBPC), 37077 G\"ottingen, Germany}
\author{F.~Javier~Garc\'{\i}a~de~Abajo}
\email{javier.garciadeabajo@nanophotonics.es}
\affiliation{ICFO-Institut de Ciencies Fotoniques, The Barcelona Institute of Science and Technology, 08860 Castelldefels (Barcelona), Spain}
\affiliation{ICREA-Instituci\'o Catalana de Recerca i Estudis Avan\c{c}ats, Passeig Llu\'{\i}s Companys 23, 08010 Barcelona, Spain}



\begin{abstract}
Spontaneous processes triggered in a sample by free electrons, such as cathodoluminescence, are commonly regarded and detected as stochastic events. Here, we supplement this picture by showing through first-principles theory that light and free-electron pulses can interfere when interacting with a nanostructure, giving rise to a modulation in the spectral distribution of the cathodoluminescence light emission that is strongly dependent on the electron wave function. Specifically, for a temporally focused electron, cathodoluminescence can be cancelled upon illumination with a spectrally modulated dimmed laser that is phase-locked relative to the electron density profile. We illustrate this idea with realistic simulations under attainable conditions in currently available ultrafast electron microscopes. We further argue that the interference between excitations produced by light and free electrons enables the manipulation of the ultrafast materials response by combining the spectral and temporal selectivity of the light with the atomic resolution of electron beams.
\end{abstract}

\maketitle 


\section{Introduction}

Coherent laser light provides a standard tool to selectively create optical excitations in atoms, molecules, and nanostructures with exquisite spectral resolution \cite{ZSC20}. Additional selectivity in the excitation process can be gained by exploiting the light polarization and the spatial distribution of the optical field to target, for example, modes with specific angular momentum in a specimen \cite{ZMD16}. However, the diffraction limit constraints our ability to selectively act on degenerate excitation modes sustained by structures that are separated by either less than half the light wavelength when using far-field optics (unless ingenious, sample-dependent schemes are adopted \cite{BPS06,Z08,YZ19}) or a few tens of nanometers when resorting to near-field enhancers such as metallic tips \cite{HTK02,BKO10,WFM14}. In contrast to light, electron beams, which are also capable of producing optical excitations \cite{paper149}, can actuate with a spatial precision roughly determined by their lateral size, currently reaching the sub-{\AA}ngstrom domain in state-of-the-art electron microscopes \cite{BDK02,KLD14,KDH19}. Indeed, the evanescent electromagnetic field accompanying a fast electron spans a broadband spectrum that mediates the transfer of energy and momentum to sample excitation modes with such degree of spatial accuracy \cite{paper149}. But unfortunately, spectral selectivity is lost because of the broadband nature of this excitation source, unless post-selection is performed by energy-filtering of the electrons, as done for instance in electron energy-loss spectroscopy (EELS) \cite{paper149,KS14,paper338}.

Photons and electrons team up to extract the best of both worlds in the rapidly evolving field of ultrafast transmission electron microscopy (UTEM), whereby the high spatial precision of electron microscopes is combined with the time resolution and spectral selectivity of optical spectroscopy. In this technique, ultrashort electron pulses created by photoelectron emission are used to track structural or electronic excitations with picosecond and femtosecond temporal resolution \cite{GLW06,BPK08,BFZ09,FZ12,PML13_2,FES15,BPC16,FBR17,HCW18,ARM20,ZZW20}. Regarding electron-photon interaction, UTEM allows us to exploit the evanescent optical field components created by light scattering at nanostructures, so that the interaction is facilitated by passing the free-electron beam through these fields, thus enabling spectrally and temporally resolved imaging with combined resolution in the nanometer--fs--meV domain {\it via} the so-called photon-induced near-field electron microscopy (PINEM) technique \cite{BFZ09,paper151,PLZ10,PZ12,KGK14,PLQ15,FES15,paper282,EFS16,KSE16,RB16,VFZ16,paper272,FBR17,PRY17,paper306,paper311,paper312,MB18,MB18_2,paper325,paper332,K19,PZG19,paper339,RML20,DNS20,KLS20,WDS20,RK20,MVG20,VMC20}. This approach has been exploited to investigate the temporal evolution of plasmons \cite{PLQ15,paper282} and optical cavity modes \cite{KLS20,WDS20}, as well as a way to manipulate the electron by exchanging transverse linear \cite{paper272,paper311,FYS20} and angular \cite{paper332,paper312} momentum with the photon field.

Following concepts from accelerator physics \cite{SCI08}, temporal compression of the electron beam into a train of attosecond pulses can be achieved by periodic momentum modulation and free-space propagation, using either ponderomotive forces \cite{BZ07,KSH18,SMY19} or PINEM-like inelastic electron-light scattering interactions \cite{FES15,PRY17,MB18,MB18_2,RTN20,MB20}. Accompanying these advances in our ability to manipulate free electrons, recent theoretical studies have explored the use of modulated free electrons to gain control over the density matrix of excitations created in a sample \cite{GY20,paperarxiv3,paper360,HRN21,ZSF21}. Intriguingly, the cathodoluminescence (CL) emission produced by a PINEM-modulated electron has been predicted to bear coherence with the laser used to achieve such modulation, which could be revealed through correlations in an interferometer \cite{paperarxiv3}. This scenario holds the potential to combine light and electrons as coherent probes acting on a sample, possibly enabling practical applications in pushing the space-time-energy levels of resolution beyond their current values. Although we refer to {\it coherence} in a precise way in what follows ({\it i.e.}, the interference of two phase-locked processes), this term can have various meanings when applied to different types of processes, so we provide a discussion of possible interpretations in the context of electron microscopy in the Supplementary Information.

The CL intensity is extremely low in most samples ($\lesssim10^{-5}$ photons per electron), unless we restrict ourselves to special classes of targets ({\it e.g.}, those enabling phase matching between the emitted radiation and the electron \cite{paper180,K19,DNS20}). When measuring far-field radiation, the visibility of the interference between CL emission and external light could be enlarged by dimming the latter to match the former. Shot noise that could potentially mask the resulting interference is avoided if photon measurements are performed at a single detector ({\it i.e.}, after the amplitudes of CL and external light have been coherently superimposed). Based on this idea, we anticipate that the use of dimmed illumination in combination with CL light emission represents a practical route towards the sought-after push in space-time-energy resolution with which we can image and manipulate optical excitations at the nanoscale.

Here, we show that the optical excitations produced in a structure by the combined effect of light and free electrons can add coherently, therefore providing a tool for actively manipulating sample excitations. The combination of light and electrons adds the spatial resolution of the latter to the spectral selectivity of the former in our ability to manipulate and probe nanoscale materials and their optical response. Specifically, we illustrate this possibility by showing that the CL emission produced by a free electron can be coherently manipulated by simultaneously exciting the sample with suitably modulated external light. We demonstrate that it is possible to strongly modulate the CL emission using currently existing technology, while complete cancellation of CL is physically feasible using tightly compressed electron wavepackets, which act as classical external point charges. The present work thus capitalizes on the correlation between CL from modulated electrons and synchronized external light as discussed in ref\ \citenum{paperarxiv3}, so we propose a disruptive form of ultrafast electron microscopy based on the direct observation of interference between CL emission and dimmed light scattering at a single photon spectrometer. We anticipate the application of interference in the excitations produced by the simultaneous action of light and electrons as a route towards spectrally resolved imaging and selective excitation of sample optical modes with an improved level of space-time-energy resolution. The sensitivity provided by the measurement of the relative phases between electron and laser waves could be further enhanced through lock-in amplification schemes that isolate the interference effects to gain information on both the electron density profile and the temporal evolution of the targeted optical excitations.

\begin{figure*}
\centering{\includegraphics[width=0.60\textwidth]{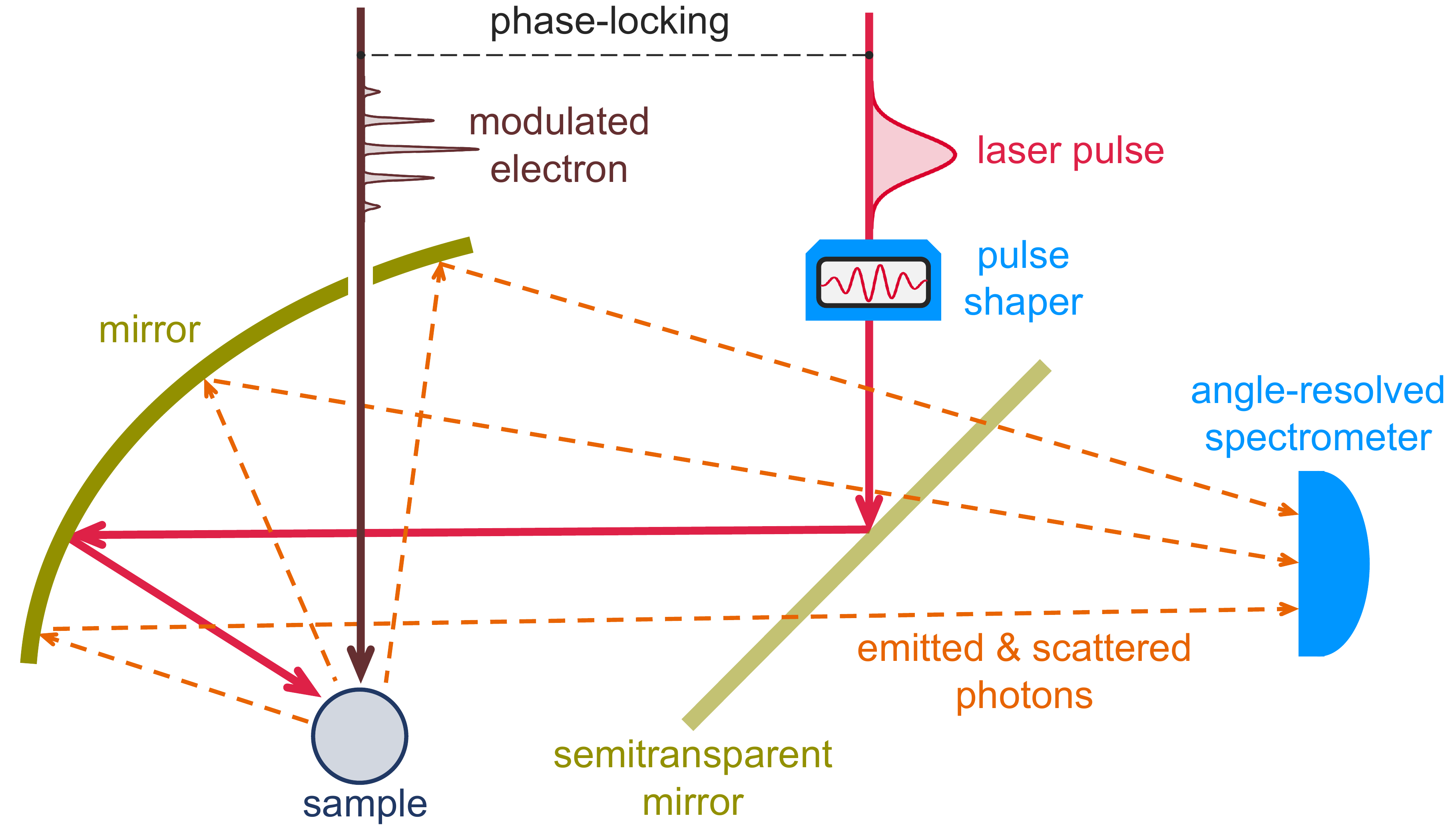}}
\caption{{\bf Sketch of the system under consideration.} A laser pulse and a modulated electron are made to interact with a sample and produce light scattering and cathodoluminescence (CL) emission, respectively. The electron is synchronized with the laser pulse to maintain mutual phase coherence. The resulting emitted and scattered photons are collected by a spectrometer. A laser pulse shaper is inserted in this scheme to bring the scattered light amplitude to a level that is commensurate with the CL emission field.}
\label{Fig1}
\end{figure*}


\section{First-Principles Description of CL Interference with External Light}

We consider the combined action of external light and free electrons on a sampled structure, such as schematically illustrated in Figure\ \ref{Fig1}. Under common conditions met in electron microscopes, the electrons can be prepared with well-defined velocity, momentum, and energy, such that their wave functions consist of components that have a narrow spread relative to those values. Additionally, we adopt the nonrecoil approximation by assuming that any interaction with the specimen produces negligible departures of the electron velocity with respect to its average value ({\it i.e.}, small momentum transfers relative to the central electron momentum). Under these conditions, we calculate the far-field radiation intensity produced by the combined contributions of interaction with the electron and scattering from a laser, based on the far-field Poynting vector. In a fully quantum treatment of radiation, the angle- and frequency-resolved far-field (ff) photon probability reduces to
\begin{align}
\frac{d\Gamma_{\rm ff}}{d\Omega_{\hat{\rb}}d\omega}
=\lim_{kr\to\infty}\;\frac{r^2}{4\pi^2\hbar k}{\rm Re}\left\{\left\langle\hat{\mathcal{E}}(\rb,\omega)\times \hat{\mathcal{B}}^\dagger(\rb,\omega) \right\rangle \right\}\cdot \hat{\rb},
\label{GEB}
\end{align}
where $k=\omega/c$ (see detailed derivation in the Appendix). This expression is the quantum counterpart of a classical result for CL \cite{paper149}, now involving the position- and frequency-dependent positive-energy part of the electric and magnetic field operators $\hat{\mathcal{E}}(\rb,\omega)$ and $\hat{\mathcal{B}}(\rb,\omega)$, respectively. We follow a quantum electrodynamics formalism in the presence of dispersive and absorptive media \cite{DKW98,paper357} to calculate this quantity for a free electron of incident wave function $\psi^0(\rb)$ and external light characterized by a spectrally resolved electric field amplitude $\Eb^{\rm ext}(\rb,\omega)$. After some analysis (see Appendix), taking the electron velocity vector $\vb$ along $z$, we find
\begin{widetext}
\begin{align}
&\frac{d\Gamma_{\rm rad}}{d\Omega_{\hat{\rb}}d\omega}=\frac{1}{4\pi^2\hbar k}\bigg{[} \int d^2\Rb'\, M_0(\Rb') |\fb_{\rr}^{\rm CL}(\Rb',\omega)|^2 \label{eqgen}\\
&\quad+|\fb_{\rr}^{\rm scat}(\omega)|^2+2\int d^2\Rb'\, {\rm Re}\left\{M_{\omega/v}(\Rb')\; \fb_{\rr}^{{\rm CL}*}(\Rb',\omega)\cdot \fb_{\rr}^{\rm scat}(\omega)\right\}\bigg{]},
\nonumber
\end{align}
\end{widetext}
where
\begin{align}
M_{\omega/v}(\Rb)=\int_{-\infty}^\infty dz \;\ee^{\ii \omega z /v }\, |\psi^0(\rb)|^2
\label{MwvR}
\end{align}
is the Fourier transform of the electron probability density, which acts as a coherence factor. Here, we use the notation $\rb=(\Rb,z)$ with $\Rb=(x,y)$ and we define the electric far-field amplitudes $\fb_{\rr}^{\rm CL}(\Rb',\omega)$ and $\fb_{\rr}^{\rm scat}(\omega)$ through the asymptotic expressions
\begin{subequations}
\label{asympfs}
\begin{align}
4\pi\ii e\omega \int dz'\,\ee^{\ii\omega z'/v}\,&G(\rb,\Rb',z',\omega)\cdot \zz \nonumber\\ &\xrightarrow[kr\to\infty]{} \frac{\ee^{\ii k r}}{r}\fb_{\rr}^{\rm CL}(\Rb',\omega),
\label{limitvectorq}\\
\Eb^{\rm scat}(\rb,\omega) &\xrightarrow[kr\to\infty]{} \frac{\ee^{\ii k r}}{r}\fb_{\rr}^{\rm scat}(\omega),
\label{limitvectorf}
\end{align}
\end{subequations}
corresponding to the classical CL and laser scattering contributions, respectively. It should be noted that we only retain the $1/r$ radiative components of the far field in $d\Gamma_{\rm rad}/d\Omega_{\hat{\rb}}d\omega$ (see eqs\ \ref{eqgen} and \ref{asympfs}), which is a legitimate procedure when considering directions in which they do not interfere with the external illumination. Nevertheless, interference between the incident and forward $1/r$ radiative components produces an additional contribution $d\Gamma_{\rm forward}/d\Omega_{\hat{\rb}}d\omega$ ({\it i.e.}, $d\Gamma_{\rm ff}/d\Omega_{\hat{\rb}}d\omega=(d\Gamma_{\rm rad}/d\Omega_{\hat{\rb}}d\omega)+(d\Gamma_{\rm forward}/d\Omega_{\hat{\rb}}d\omega)$), as we discuss below in relation to the energy pathways associated with the interaction. The specimen is assumed to be characterized by a linear and local electromagnetic response, which enters this formalism through the Green tensor, implicitly defined by
\begin{align}
\nabla\times\nabla\times G(\rb,\rb',\omega)&-k^2\epsilon(\rb,\omega)G(\rb,\rb',\omega) \nonumber\\
&=-\frac{1}{c^2}\delta(\rb-\rb'),
\label{greentensor}
\end{align}
where $\epsilon(\rb,\omega)$ is the position- and frequency-dependent permittivity. The first and second terms in eq\ \ref{eqgen} describe the separate contributions from CL and light scattering, respectively, whereas the third term accounts for interference between them. We remark that this result relies on the nonrecoil approximation for the electron, which allows us to replace its associated current operator by the average expectation value under the the assumption that $\vb$ remains unaffected by the interaction.

Interestingly, the CL emission in the absence of external illumination ({\it i.e.}, the first term in eq\ \ref{eqgen}), is constructed as an incoherent sum of contributions from different lateral positions $\Rb'$ across the electron beam \cite{RH1988,paperarxiv2} ({\it i.e.}, no interference remains in this signal between the CL emission from different lateral positions of the beam). In contrast, the signal associated with the interference between CL and light scattering (third term in eq\ \ref{eqgen}) contains further interference between the contribution of different lateral electron-beam positions $\Rb'$. Interestingly, this effect is genuinely associated with interference between different lateral positions of the beam because the light scattering amplitude $\fb_{\rr}^{\rm scat}(\omega)$ in that equation does not depend on $\Rb'$.

For completeness, we note that eq\ \ref{eqgen} can be written in the more compact form
\begin{align}
&\frac{d\Gamma_{\rm rad}}{d\Omega_{\hat{\rb}}d\omega} =\frac{1}{4\pi^2\hbar k}\nonumber\\
&\times\int d^3\rb'\,|\psi^0(\rb')|^2
\big{|}\ee^{-\ii\omega z'/v}\,\fb_{\rr}^{\rm CL}(\Rb',\omega)+\fb_{\rr}^{\rm scat}(\omega)\big{|}^2 \nonumber,
\end{align}
which directly reflects the interference between CL and laser scattering. In addition, our results can easily be generalized to deal with several distinguishable electrons (labeled by superscripts $j$), for which we have
\begin{widetext}
\begin{align}
&\frac{d\Gamma_{\rm rad}}{d\Omega_{\hat{\rb}}d\omega}=\frac{1}{4\pi^2\hbar k}\bigg{\{} \sum_j\int d^2\Rb'\, M_0^j(\Rb') |\fb_{\rr}^{\rm CL}(\Rb',\omega)|^2 \label{multielectrons}\\
&\quad+|\fb_{\rr}^{\rm scat}(\omega)|^2+2\sum_j\int d^2\Rb'\, {\rm Re}\left\{M_{\omega/v}^j(\Rb')\; \fb_{\rr}^{{\rm CL}*}(\Rb',\omega)\cdot \fb_{\rr}^{\rm scat}(\omega)\right\} \nonumber\\
&+\sum_{j\neq j'}
\left[\int d^2\Rb'\, M_{\omega/v}^j(\Rb')\fb_{\rr}^{{\rm CL}*}(\Rb',\omega)\right]
\left[\int d^2\Rb''\, M_{\omega/v}^{j'*}(\Rb'')\fb_{\rr}^{\rm CL}(\Rb'',\omega)\right]
\bigg{\}}
\nonumber
\end{align}
\end{widetext}
(see derivation in the Appendix), where $M^j_{\omega/v}$ is given by eq\ \ref{MwvR} with $\psi^0$ replaced by $\psi^j$ (the wave function of electron $j$). In the absence of external light ({\it i.e.}, with $\fb_{\rr}^{\rm CL}=0$), this expression converges to the multi-electron excitation probability described elsewhere \cite{paperarxiv2}.

\indent While the above results are derived for electrons prepared in pure states (i.e., with well-defined wave functions), the extension to mixed electron states is readily obtained by evaluating the averages in eqs\, (\ref{avcurrent}) as ${\rm Tr}\{\hat{\jb}^{\rm el}(\rb',\omega)\hat{\jb}^{\rm el}(\rb'',\omega)\hat{\rho}^j\}$ and ${\rm Tr}\{\hat{\jb}^{\rm el}(\rb',\omega)\hat{\rho}^j\}$, respectively, where $\hat{\rho}^j$ is the electron density matrix of electron $j$. This leads exactly to the same expressions as above but replacing $|\psi^j(\rb)|^2$  by the probability densities $\langle \rb |\hat\rho^j|\rb \rangle$, which allow us to describe electrons that have undergone decoherence processes before interacting with the sample.

\indent We present results below for nanoparticles whose optical response can be described through an isotropic, frequency-dependent polarizability $\alpha(\omega)$. Considering a well-focused electron with impact parameter $\Rb_0$ relative to the particle position $\rb=0$ ({\it i.e.}, an electron probability density $|\psi^0(\rb)|^2\approx\delta(\Rb-\Rb_0)|\psi_\parallel(z)|^2$), we find that eq\ \ref{eqgen} then reduces to
\begin{align}
&\frac{d\Gamma_{\rm rad}(\Rb_0)}{d\omega}=\frac{2k^3}{3\pi\hbar} |\alpha(\omega)|^2 \nonumber\\
&\times\bigg[\left|\Eb^{{\rm ext}}(0,\omega)+M_{\omega/v}^*\,\Eb^{\rm el}(\Rb_0,\omega)\right|^2
\nonumber\\
&\quad\quad+\left(1-\left|M_{\omega/v}\right|^2\right)\,|\Eb^{\rm el}(\Rb_0,\omega)|^2 \bigg],
\label{CLdip}
\end{align}
where
\begin{align}
&\Eb^{\rm el}(\Rb_0,\omega)
=\frac{2e\omega}{v^2\gamma}\,\left[K_1\left(\frac{\omega R_0}{v\gamma}\right)\,\RR_0+\frac{\ii}{\gamma}K_0\left(\frac{\omega R_0}{v\gamma}\right)\,\zz\right],
\label{FF}
\end{align}
$\gamma=1/\sqrt{1-v^2/c^2}$ is the Lorentz factor, and we now have
\begin{align}
M_{\omega/v}=\int_{-\infty}^\infty dz \;\ee^{\ii \omega z /v }\, |\psi_\parallel(z)|^2
\label{Mwv}
\end{align}
for the electron coherence factor. These expressions clearly reveal that, although the phase of the electron wave function is erased because only the probability density appears in eq\ \ref{Mwv}, the mutual electron-light coherence is controlled by the temporal profile of that density, as well as its timing with respect to the light field, which produces a global phase in $M_{\omega/v}$ relative to the light field that in turn enters through the first term inside the square brackets in eq\ \ref{CLdip} ({\it e.g.}, to partially cancell the CL emission). Obviously, without electron-laser timing, averaging over this phase difference cancels such interference.

Reassuringly, eq\ \ref{CLdip} reduces to well-known expressions for the CL emission when setting $\Eb^{\rm ext}=0$ ({\it i.e.}, in the absence of external light). This result is independent of the electron wave function \cite{PG19,paperarxiv3,paper360,paperarxiv2}. Moreover, we recover the photon scattering cross section $\propto\omega^3|\alpha|^2$ when $\Eb^{\rm el}=0$ ({\it i.e.}, without the electron). An additional element of intuition is added by the fact that the expression for $\Eb^{\rm el}(\Rb_0,\omega)$ corresponds to the spectrally resolved evanescent field produced by a classical point electron \cite{paper149}, which decays exponentially away from the trajectory, as described by the modified Bessel functions $K_0$ and $K_1$.

The electron coherence factor $M_{\omega/v}$ in eq\ \ref{Mwv} (and similarly $M_{\omega/v}(\Rb)$ in eq\ \ref{MwvR}) determines the degree of coherence (DOC) of the electron excitation ({\it i.e.}, the CL emission) relative to the signal originating in the laser ({\it i.e.}, light scattering). This factor enters eq\ \ref{CLdip} through terms proportional to ${\rm DOC}(\omega)=|M_{\omega/v}|^2$, where we use the definition of DOC introduced in ref\ \citenum{paperarxiv3}. Indeed, for $M_{\omega/v}=0$, the scattered light field does not mix at all with the CL emission field, so they are mutually incoherent. In contrast, if $M_{\omega/v}=1$, we have a maximum of coherence, so that the external illumination can fully suppress the CL emission. Specifically, we stress that the point-particle limit of the electron ({\it i.e.}, $|\psi^0(\rb)|^2\rightarrow\delta(\rb)$) produces $M_{\omega/v}=1$, thus recovering the intuitive result for a classical point charge: the radiation from the passage of the electron is a deterministic solution of the Maxwell equations, and thus, it can be suppressed by an external light field with the same frequency-dependent amplitude and an opposite phase. This is not the case in general, so for arbitrarily distributed electron wave functions, the degree of coherence is partially reduced. We also note that the phase of the electron wave function is entirely removed from the coherence factor (see eq\ \ref{MwvR}).

We have shown that the CL emission can be modulated by interference with external laser light. As a way to illustrate this effect, we discuss in what follows the maximum achievable minimization of the overall far-field (scattered+emitted) photon intensity by appropriately selecting the external far-field amplitude. If we have complete freedom to choose the external field, we readily find from eq\ \ref{CLdip} that $d\Gamma_{\rm rad}/d\omega$ is minimized by taking
\begin{align}
\Eb^{\rm ext}(0,\omega)=-M^*_{\omega/v} \Eb^{{\rm el}}(\Rb_0,\omega).
\label{optimum}
\end{align}
Alternatively, when one adopts light pulses $\Eb^{\rm ext}(0,\omega)=f(\omega)\,\Eb_0$ with a predetermined spectral profile $f(\omega)$ ({\it e.g.}, a Gaussian $f(\omega)=\ee^{-(\omega-\omega_0)^2\sigma_t^2/2}$), the minimization condition at a given sample resonance frequency $\omega=\omega_0$ is readily achieved by setting the field amplitude to $\Eb_0=-M_{\omega_0/v}^*\Eb^{{\rm el}}(\Rb_0,\omega_0)\,f^*(\omega_0)/|f(\omega_0)|^2$. As an estimate of the laser intensity needed to optimally modulate the CL emission, we take $\big|M_{\omega/v}\big|=1$ and consider the electric field amplitude from eq {\blue 8} for a 100\,keV electron passing at a distance $R_0=$50\,nm (10\,nm) away from the dipolar particle, so that, setting $\hbar\omega=1\,$eV, we have $|\Eb^{\rm el}(\Rb_0,\omega)|\Delta\omega\sim50\,$kV/m (280\,kV/m), assuming a depletion bandwidth $\hbar\Delta\omega=0.1\,$eV; also, the corresponding laser fluence is $(c/4\pi^2)|\Eb^{\rm el}(\Rb_0,\omega)|^2\Delta\omega\sim10\,$nJ/m$^2$ (400\,nJ/m$^2$).

\begin{figure*}
\centering{\includegraphics[width=0.70\textwidth]{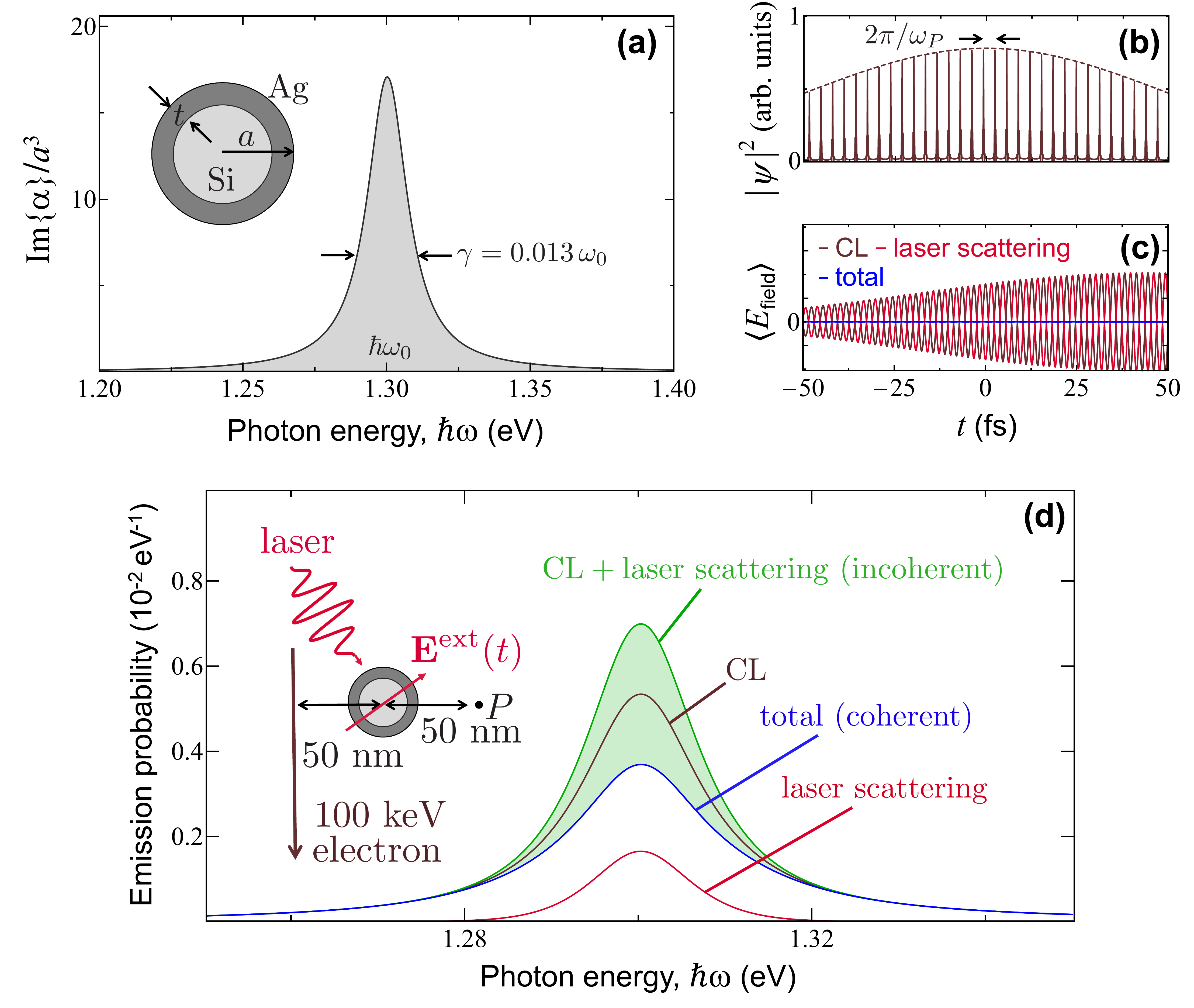}}
\caption{{\bf Interference between cathodoluminescence and external light scattering.} (a) We consider a sample consisting of a small isotropic scatterer described through a frequency-dependent polarizability $\alpha(\omega)$ that is dominated by a single resonance of frequency $\omega_0$ and width $\gamma$. For concreteness, we take a nanosphere (see inset) comprising a silicon core (60\,nm diameter, $\epsilon=12$ permittivity) coated with a silver layer (5\,nm thickness, permittivity taken from optical data \cite{JC1972}), for which $\hbar\omega_0=1.3\,$eV and $\gamma=0.013\,\omega_0$. In the plot, the polarizability is normalized using the outer particle radius $a=35\,$nm. (b) Electron density profile of a 100\,keV electron Gaussian wavepacket (50\,fs standard-deviation duration in probability density) after modulation through PINEM interaction (coupling coefficient $|\beta|=5$, central laser frequency tuned to $\omega_P=\omega_0$) followed by free propagation over a distance $d=2.5$\,mm, which produces a train of temporally compressed density pulses. (c) Time dependence of the CL, laser-scattering, and total field amplitudes for the electron in (b) and a laser Gaussian pulse of 50\,fs duration in amplitude. The light amplitude is optimized to deplete the CL signal at frequency $\omega_0$. (d) Spectral dependence of the resulting angle-integrated far-field CL (maroon curve), laser-scattering (red curve), and total (blue curve) light intensity for the optimized amplitude of the Gaussian laser pulse. The incoherent sum of CL emission and laser scattering signals is shown for comparison (green curves). The shaded region corresponds to spectra obtained with partially optimized laser pulses. The inset in (d) shows details of the geometry under consideration, also indicating the position $P$ at which the field in (c) is calculated.}
\label{Fig2}
\end{figure*}

\section{Results and Discussion}

Motivated by potential application of electron beams in controlling the excitations of small elements in a sample ({\it e.g.}, molecules), we consider a dipolar scatterer as that depicted in Figure\ \ref{Fig2}a, consisting of a 60\,nm silicon sphere coated with a silver layer of 5\,nm thickness ({\it i.e.}, an outer radius $a=35\,$nm), which exhibits a spectrally isolated plasmon resonance at a photon energy $\hbar\omega_0=1.3\,$eV. In practice, we calculate the dipolar polarizability of small spheres from the corresponding electric Mie scattering coefficient as $\alpha=(3/2k^3)t_1^E$ \cite{paper149}. The relatively low level of ohmic losses in silver produces a narrow resonance, with 14\% of its FWHM ($\hbar\gamma=0.013\,\hbar\omega_0\approx17\,$meV) attributed to radiative losses, as estimated from the ratio ($\approx0.86$) of peak absorption to extinction cross sections. Similar dipolar resonances can be found in other types of samples, such as metallic nanoparticles of different morphology \cite{GPM08,LM13} and dielectric cavities \cite{paperxx3}, for which we anticipate a variability in their coupling strength to light and electrons that should not however affect the qualitative conclusions of the present work.

In what follows, we consider modulated electrons, focusing on their interaction with a particle under simultaneous laser irradiation. The production of sub-fs-modulated electrons has become practical thanks to PINEM-related advances in ultrafast electron microscopy, whereby an ultrashort laser pulse is used to mould each electron into a train of pulses \cite{KML17,PRY17,MB18_2,MB18,SMY19,RTN20}, from which an individual wavepacket can be extracted by applying a streaking technique \cite{MB20}. Specifically, we consider either Gaussian electron wavepackets defined by the wave function
\begin{align}
\psi_\parallel(z)=\frac{1}{(2\pi\sigma_t^2v^2)^{1/4}}\ee^{-z^2/4\sigma_t^2v^2+\ii q_0z},
\label{Gaussian}
\end{align}
where the duration is expressed in terms of the standard deviation $\sigma_t$ of the electron pulse probability density $|\psi_\parallel(z)|^2$ and $q_0$ is the central wave vector, or electrons modulated by PINEM interaction with scattered laser light followed by free-space propagation over a macroscopic distance $d$ before reaching the sampled particle. The wave function of the so modulated electron consists of a Gaussian wavepacket envelope ({\it i.e.},  eq\ \ref{Gaussian}) multiplied by an overall modulation factor \cite{paper360,paperarxiv2}
\begin{align}
\mathcal{P}_d(\beta,\omega,z)
=\sum_{l=-\infty}^\infty J_l(2|\beta|)\,\ee^{\ii l\omega_P(z-z_P)/v-2\pi\ii l^2d/z_T},
\label{PPINEM}
\end{align}
where $l$ labels a periodic array of energy sidebands separated by the laser photon energy $\hbar\omega_P$; the modulation strength is quantified by a single complex coupling parameter $\beta$  that is proportional to the laser amplitude and whose phase determines the reference position $z_P$; and we have introduced a sideband-dependent recoil correction phase $\propto l^2$ to account for propagation over $d$, involving a Talbot distance $z_T=4\pi\me v^3\gamma^3/\hbar\omega_P^2$. These expressions are valid under the assumption that the laser is quasi-monochromatic ({\it i.e.}, its frequency spread is small compared with $\omega_P$). Then, for an optimum value of $d$, the factor $\mathcal{P}_d(\beta,\omega,z)$ renders a temporal comb of periodically spaced pulses (time period $2\pi/\omega_P$) that are increasingly compressed as $|\beta|$ is made larger, eventually reaching attosecond duration \cite{KML17,PRY17,MB18_2,MB18,SMY19,RTN20}. We remark that mutual electron-laser phase coherence can be achieved by using the same laser to both modulate the electron and subsequently interact with the sample. For concreteness, we set the electron energy to 100\,keV and tune the PINEM laser frequency to the resonance of the aforementioned sample ({\it i.e.}, $\hbar\omega_P=\hbar\omega_0=1.3\,$eV). The corresponding Talbot distance is then $z_T\approx211$\,mm.

\subsection{Optical Modulation of CL from a Dipolar Scatterer}

An example of PINEM-modulated electron density profile is shown in Figure\ \ref{Fig2}b for $\sigma_t=50\,$fs, $|\beta|=5$, and $d=2.5\,$mm. Direct application of eq\ \ref{CLdip} to this electron allows us to calculate the CL emission spectrum, along with its modulation due to interference with light scattering from a phase-locked Gaussian pulse (50\,fs duration in field amplitude), as shown in Figure\ \ref{Fig2}d, where the inset depicts further details of the geometrical arrangement and configuration parameters. Starting from the CL spectrum in the absence of external illumination (maroon curve, which we insist is independent of electron wave function profile \cite{PG19,paperarxiv3,paper360,paperarxiv2}), we then superimpose the phase-locked laser pulse in which we optimize the light field amplitude $\Eb_0$ as prescribed above to produce a maximum of depletion in the resulting photon intensity at the peak maximum (blue curve). The achievable depletion is not complete because we have ${\rm DOC}(\omega_0)=|M_{\omega_0/v}|^2\approx0.31$ for the considered electron, which differs from the limit of perfect coherence (see below), so a fraction of the original CL signal given by $1-{\rm DOC}(\omega_0)\approx69\%$ remains after complete cancellation of the coherent part. If the electron and light pulses are not phase-locked, relative phase averaging renders $M_{\omega/v}=0$, so the resulting probability of detecting CL or scattered photons (green curve) is just the incoherent sum of the probabilities associated with these two processes ({\it i.e.}, the sum of the blue and red curves).

It is instructive to compare the electric near field associated with CL {\it versus} light scattering by computing the quantum average of the corresponding field operator $\hat{\Eb}^{\rm H}(\rb,t)$. Although this quantity is an observable, we note that its measurement is not straightforward. Following the approach explained in the Appendix and retaining only terms that are linear in the electron current operator $\hat{\jb}(\rb,\omega)$, we find the average field to be given by $\langle \hat{\Eb}^{\rm H}(\rb,t) \rangle=-2 \ii \int_{-\infty}^{\infty} \omega\,d\omega\, \ee^{-\ii \omega t} \int d^3\rb'\, G(\rb,\rb',\omega) \cdot \langle\hat{\jb}(\rb^{\prime},\omega)\rangle$, which under laser and electron exposure becomes
\begin{align}
&\left\langle \hat{\Eb}^{\rm H}(\rb,t)\right\rangle=(2\pi)^{-1}\int_{-\infty }^\infty d\omega \,\ee^{-\ii \omega t} \nonumber\\
&\times\left[\Eb^{\rm light}(\rb,\omega)+\int d^2\Rb'\, M^*_{\omega/v}(\Rb')\,\Eb^{\rm CL}(\rb,\Rb',\omega)\right],
\nonumber
\end{align}
where $\Eb^{\rm CL}$ is defined in the Appendix (eq\ \ref{ECL}). The scattered part of the resulting time-dependent field is plotted in Figure\ \ref{Fig2}c as calculated from this equation at the position $P$ indicated in the inset of Figure\ \ref{Fig2}d. We corroborate that the optimized laser scattering field (red) can be made to cancel the CL field (blue), therefore producing a nearly vanishing total field (blue) that is consistent with the depletion of CL observed in Figure\ \ref{Fig2}d. It is important to stress that the average of the electric field amplitude cancels, while non-vanishing fluctuations give rise to the incoherent part of the emission, which is not suppressed.

\begin{figure*}
\centering{\includegraphics[width=0.85\textwidth]{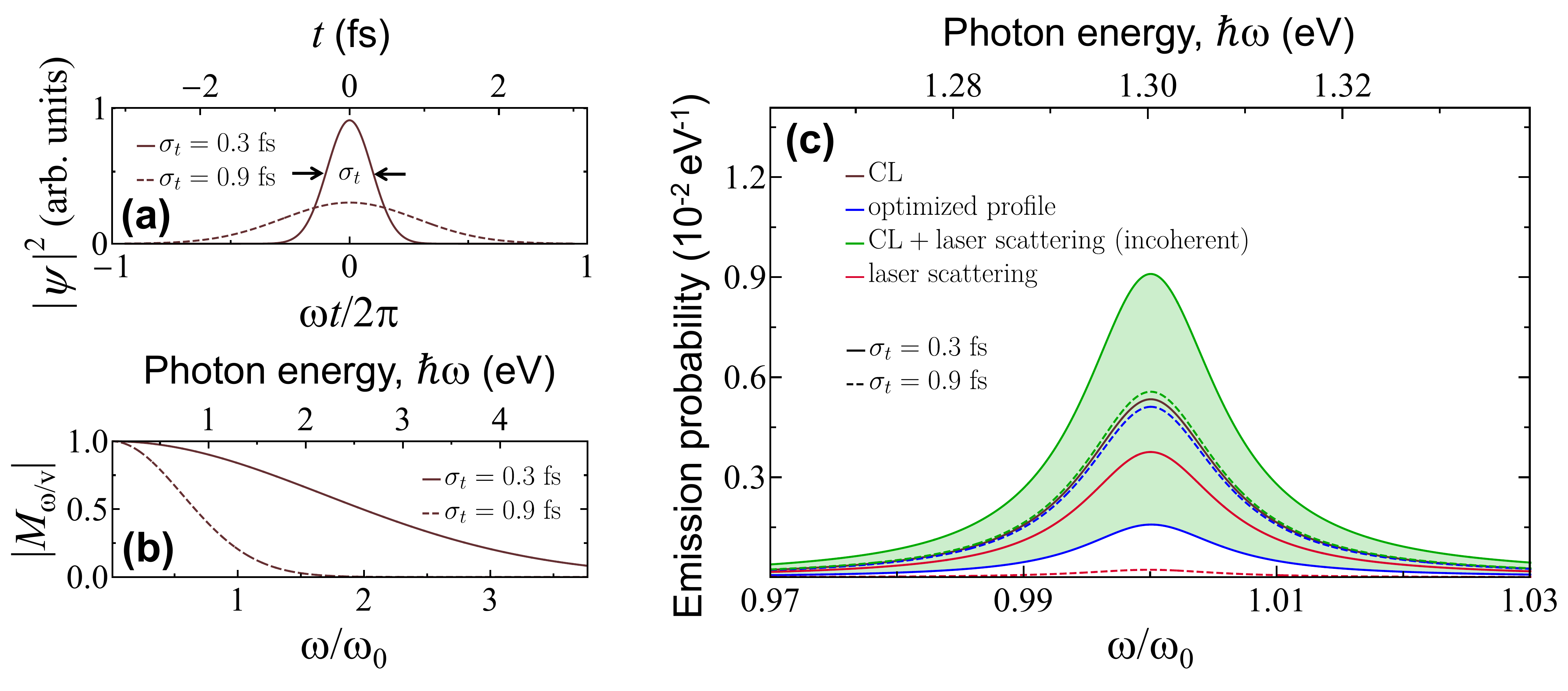}}
\caption{{\bf Modulation of the CL emission by Gaussian electron and laser pulses.} (a) Gaussian electron wavepackets of 0.3\,fs and 0.9\,fs duration. (b) Frequency dependence of the electron coherence factor $M_{\omega/v}$ (Fourier transform of the profiles in (a)). (c) Angle-integrated CL, laser scattering, and total far-field photon intensity using the electron pulses in (a), the same particle and geometrical configuration as in Figure\ \ref{Fig2}, and an optimized spectral profile of laser field amplitude. We also show the incoherent sum of CL emission and laser scattering signals for comparison (green curve).}
\label{Fig3}
\end{figure*}

\subsection{CL Modulation for Gaussian Electrons}

In Figure\ \ref{Fig3}, we consider an electron prepared in a Gaussian wavepacket with standard-deviation duration $\sigma_t$ of either 0.3\,fs or 0.9\,fs (Figure\ \ref{Fig3}a). These values are consistent with those achieved in recent experiments \cite{MB20}. The corresponding coherence factor $M_{\omega/v}=\ee^{-\omega^2\sigma_t^2/2}$ (Figure\ \ref{Fig3}b, as calculated from eqs\ \ref{Mwv} and \ref{Gaussian}) quickly dies off as the electron pulse duration exceeds the optical period $2\pi/\omega$ of the targeted excitation. In the point-electron limit ($\sigma_t\rightarrow0$), full coherence is obtained in accordance with the intuitive picture that the electron then generates a classical field that is well described by the solution of Maxwell's equations for a classical external source. The corresponding CL emission probability (Figure\ \ref{Fig3}c, maroon curve) is again independent of electron wave function, while maximal depletion can be obtained upon sample irradiation with an optimum spectral profile of the external field amplitude (eq\ \ref{optimum}), so that only a fraction $1-|M_{\omega/v}|^2$ of the CL emission remains (see eq\ \ref{CLdip}). Consequently, the level of depletion depends dramatically on pulse duration, as illustrated by comparing solid and dashed curves in Figure\ \ref{Fig3}c.

\begin{figure*}
\centering{\includegraphics[width=0.80\textwidth]{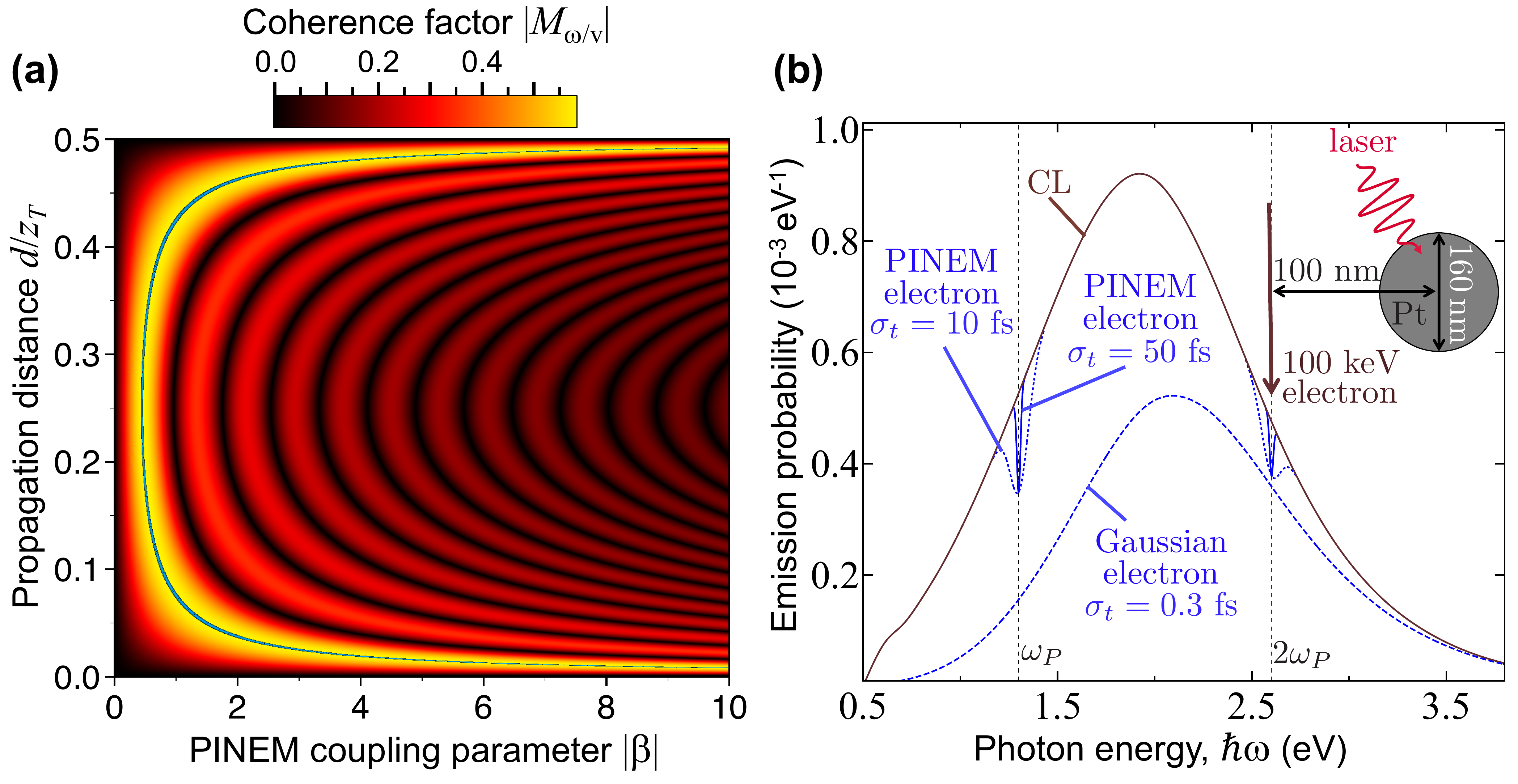}}
\caption{{\bf Coherence factor of PINEM-modulated electrons.} (a) We show the coherence factor $|M_{\omega/v}|$ for modulated electrons in the limit of long pulse duration ($\omega\sigma_t\gg1$) as a function of the PINEM coupling parameter $\beta$ and free propagation distance $d$. This function is periodic along $d$ with a period given by half the Talbot distance $z_T$. Additionally, $|M_{\omega/v}|$ presents an absolute maximum of $\approx0.582$ along the blue contour superimposed on the density plot. (b) Unperturbed (maroon curve) and optically depleted (blue curves) CL spectra from a 160\,nm Pt spherical particle and electrons prepared in Gaussian wavepacket (dashed blue curve, $\sigma_t=0.3$\,fs) or PINEM-modulated (solid and dotted blue curves obtained with $|\beta|=5$ and either $\sigma_t=50\,$fs or $\sigma_t=10\,$fs, see labels) states. The inset shows the geometrical arrangement and parameters. The laser amplitude is taken to be optimized for all emission frequencies.}
\label{Fig4}
\end{figure*}

\subsection{CL Modulation for PINEM-Compressed Electrons}

The wave function of a PINEM-modulated electron at the sample interaction region is given by the product of eqs\ \ref{Gaussian} and \ref{PPINEM} when using a quasi-monochromatic laser. The corresponding coherence factor, calculated from eq\ \ref{Mwv} as explained in the Appendix, is a function of the PINEM coupling coefficient $\beta$, the free propagation distance $d$, the excitation frequency $\omega$, the electron velocity $v$, and a slowly varying envelope profile that we take here to be a Gaussian of temporal width $\sigma_t$. In the $\omega\sigma_t\gg1$ limit, which is reached in practice with $\sigma_t\sim2\,$fs for sample excitations of $\hbar\omega=1.3\,$eV energy (see supplementary Figure\ S1), we obtain the universal plot for $|M_{\omega/v}|$ shown in Figure\ \ref{Fig4}a, where the dependence on $\omega$, $d$, and $v$ is fully encapsulated in the $d/z_T$ ratio, using the Talbot distance $z_T$ defined above. Importantly, we find a region of maximum coherence (blue contour) in which $|M_{\omega/v}|\approx0.582$, and therefore, the fraction of excitations produced by the electron that are coherent with respect to the external phase-locked laser is limited to ${\rm DOC}(\omega)=|M_{\omega/v}|^2\le34\%$ . This maximum value can be reached for coupling parameters $|\beta|\ge0.46$, while the corresponding free-propagation distance $d$ can be controlled by changing the modulating laser intensity. We note that the $d$ position at which maximum coherence is found does not coincide with that of maximal temporal compression of the electron pulse train due to a substantial electron probability density remaining in the region between consecutive peaks \cite{paper360}.

In Figure\ \ref{Fig4}b, we consider a dipolar scatter with a broad spectral response to better illustrate the optically driven depletion of CL for PINEM-compressed electrons. In particular, we take a 160\,nm Pt spherical particle, which produces a wide CL emission peak (maroon curve). For comparison, we show the depletion obtained under optimized laser irradiation ({\it i.e.}, with the external light field amplitude given in eq\ \ref{optimum}) for a Gaussian electron wavepacket of 0.3\,fs duration (Figure\ \ref{Fig4}b, dashed curve), showing a stronger effect at lower photon energies in accordance with Figure\ \ref{Fig3}b. In contrast, for a nearly-optimum PINEM-modulated electron (the same as in Figure\ \ref{Fig2}b), we find instead discrete depletion features, corresponding to the PINEM energy $\omega_P$ ({\it i.e.}, $\hbar\omega_P=1.3\,$eV in this case) and its harmonics $\omega=m\omega_P$ (only $m=1$ and 2 peaks are visible in the solid and dotted curves of Figure\ \ref{Fig4}b). We note that the leftmost depletion does not reach as deep as that produced by the Gaussian wavepacket electron, whereas the second one has nearly the same magnitude. In the $\omega\sigma_t\gg1$ limit, the depletion observed at the excitation frequencies $\omega=m\omega_P$ is equally ruled by universal plots of $M_{\omega/v}$ analogous to that in Figure\ \ref{Fig4}a (see supplementary Figure\ S2), showing a similar dependence on $\beta$ and $d$, but with an increasingly reduced magnitude as the harmonic order $m$ is increased. When the envelope of the PINEM-modulated electron is reduced from 50\,fs (solid blue curved) to 10\,fs (dotted curve), the depletion features are broadened, but their depth is maintained, directly mimicking the behavior of ${\rm DOC}(\omega)$. In other words, shorter electron pulses allow us to suppress a larger fraction of the CL power, and of course, this suppression requires illuminating the sample with a synchronized, amplitude-optimized laser that covers the range of sampled excitation frequencies $\omega$.

\begin{figure*}
\centering{\includegraphics[width=1.00\textwidth]{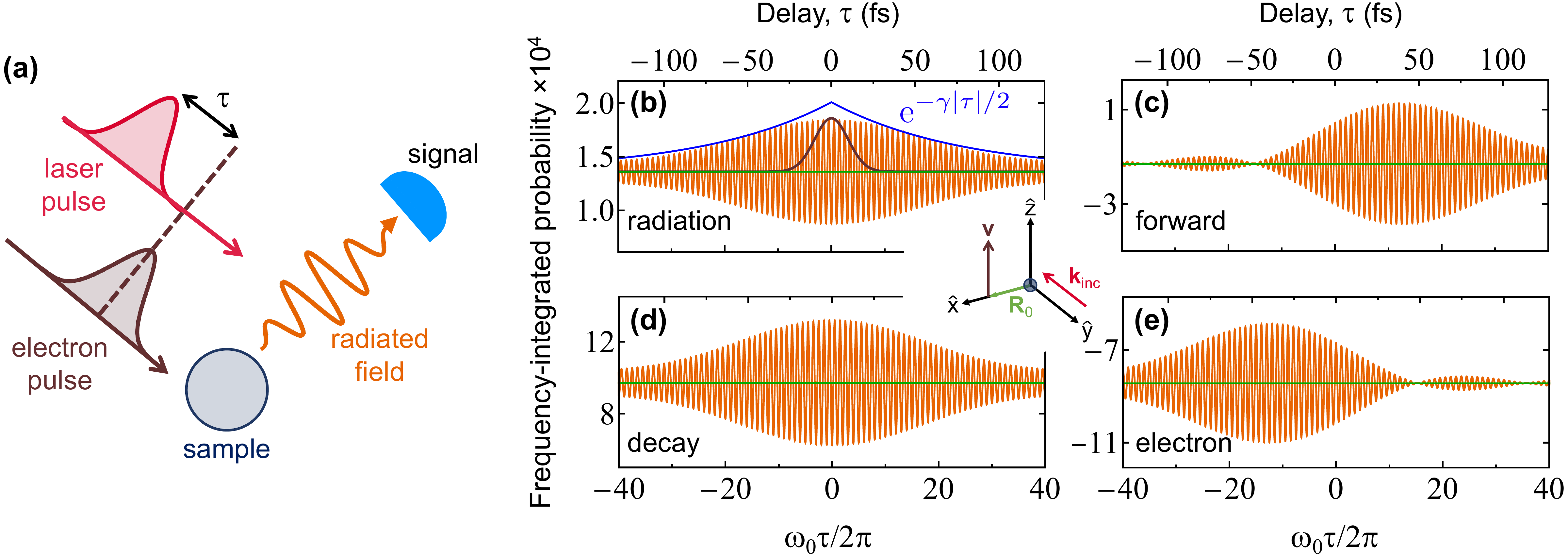}}
\caption{{\bf Control of the far-field photon intensity and energy pathways through the electron-laser temporal delay.} (a) We consider the same configuration as in Figure\ \ref{Fig2}, using electron and laser Gaussian pulses that act on the sample with a relative time delay $\tau$. (b) Angle- and frequency-integrated photon intensity (orange, in units of photons per electron), showing oscillations of period $2\pi/\omega_0$ as a function of $\tau$, as calculated for 100\,keV electrons, 10\,fs Gaussian pulse durations ({\it i.e.}, $f(\omega)=\ee^{-(\omega-\omega_0)^2\sigma_t^2/2}$, see black profile for comparison, referring to the standard deviation of the electron density profile and the laser field amplitude), and the same particle as in Figure\ \ref{Fig2}a. The interference attenuation for $\gamma\tau\gg1$ is indicated by the blue curve, where $\gamma$ is the decay rate of the sampled resonance. The laser field amplitude is fixed to $(1.4\,e\omega_0/v^2\gamma)\,f(\omega)$ and contained within the plane of the figure. (c-e) Frequency-integrated probability associated with additional energy pathways: laser-stimulated forward scattering (c), total decay following excitation of the particle plasmon (d), and change in the electron energy (e). Calculations in (b-e) correspond to the orientations of the light (incident wave vector $\kb_{\rm inc}$) and the electron (velocity $\vb$) shown in the central inset.}
\label{Fig5}
\end{figure*}

\subsection{Temporal Control of the Emission}

The studied CL modulation strongly depends on the timing between the laser and electron interactions with the sampled structure, as illustrated in Figure\ \ref{Fig5}. To elaborate on this point, we reduce the number of parameters by considering electron wavepackets with a Gaussian profile ({\it i.e.}, without an additional PINEM modulation) and vary their temporal delay relative to laser pulses (see sketch in Figure\ \ref{Fig5}a), using the same standard deviation duration $\sigma_t=10\,$fs both for the electron probability density and for the light field amplitude. We  consider the same particle as in Figure\ \ref{Fig1} and integrate the CL signal over frequency to cover the resonance region. The result is plotted in  Figure\ \ref{Fig5}b. For optimal CL suppression, the polarization induced in the particle by the electron and the laser must have overlapping envelopes with a temporal delay precision well below an optical cycle. For finite delay, we show that the interference signal oscillates as a function of $\tau$ with a period that coincides with the resonance optical period $2\pi/\omega_0$. Additionally, the amplitude of these oscillations is effectively attenuated by a factor $\ee^{-\gamma|\tau|/2}$ away from zero delay; this attenuation takes place at a pace that is half of the resonance decay rate $\gamma$ because interference is governed by the resonance amplitude rather than the intensity.

\subsection{Energy Pathways}

We present an alternative density-matrix formalism in the Supplementary Information to describe the combined electron and light interaction with an isotropic dipolar sample that hosts a triply degenerate optical mode of frequency $\omega_0$. This allows us to obtain partial probabilities for processes associated with energy changes in the electron ($\Gamma_{\rm el}$), accumulated excitations and subsequent decays of the particle mode ($\Gamma_{\rm decay}$), emission of radiation along forward ($\Gamma_{\rm forward}$) and non-forward ($\Gamma_{\rm rad}$) directions, and inelastic absorption events ($\Gamma_{\rm abs}$). This analysis leads to the following expressions for the corresponding frequency-resolved probabilities:
\begin{widetext}
\begin{subequations}
\label{allgammas}
\begin{align}
&\frac{d\Gamma_{\rm el}}{d\omega}=
-\frac{1}{\pi\hbar}\,\Imm\bigg\{
\alpha(\omega)\,\Eb^{{\rm ext}}(0,\omega)\cdot\Eb^{{\rm el}*}(\Rb_0,\omega)\,M_{\omega/v}\bigg\}
-\frac{1}{\pi\hbar}
\,\left|\Eb^{\rm el}(\Rb_0,\omega)\right|^2\,\Imm\left\{\alpha(\omega)\right\}, \\ 
&\frac{d\Gamma_{\rm decay}}{d\omega}=\frac{1}{\pi\hbar}\,
\Imm\{\alpha(\omega)\}\;\bigg[
|\Eb^{\rm ext}(0,\omega)|^2+|\Eb^{\rm el}(\Rb_0,\omega)|^2
+2\,\Ree\big\{\Eb^{\rm ext}(0,\omega)\cdot\Eb^{{\rm el}*}(\Rb_0,\omega)M_{\omega/v}\big\}\bigg], \\ 
&\frac{d\Gamma_{\rm forward}}{d\omega}
=-\frac{1}{\pi\hbar}\,\Imm\big\{\alpha(\omega)\,\Eb^{{\rm ext}*}(0,\omega)\cdot
[\Eb^{\rm ext}(0,\omega)+\Eb^{\rm el}(\Rb_0,\omega)M_{\omega/v}^*]\big\}.  \label{forward} 
\end{align}
\end{subequations}
\end{widetext}
In addition, it reproduces eq\ \ref{CLdip} for $d\Gamma_{\rm rad}/d\omega$, whereas the remaining absorption probability is given by $d\Gamma_{\rm abs}/d\omega=(d\Gamma_{\rm decay}/d\omega)-(d\Gamma_{\rm rad}/d\omega)$. Importantly, the probabilities in eqs\ \ref{allgammas} satisfy the energy-conservation condition
\begin{align}
\frac{d\Gamma_{\rm el}}{d\omega}+\frac{d\Gamma_{\rm decay}}{d\omega}+\frac{d\Gamma_{\rm forward}}{d\omega}=0.
\label{balance}
\end{align}
To corroborate the correctness of these results, we have obtained an independent derivation of eqs\ \ref{allgammas} based on an extension of the quantum-electrodynamics formalism followed in the Appendix, as succinctly described in the Supplementary Information.

We interpret $\Gamma_{\rm forward}$ as the change in photon forward emission ({\it i.e.}, toward the direction of propagation of the incident light beam) associated with interference between emitted and externally incident photons ({\it i.e.}, the type of stimulated process that is neglected in the non-forward far-field radiation probability $\Gamma_{\rm rad}$). In particular, the first term inside the squared brackets of eq\ \ref{forward} agrees with the depletion of the incident light that is described by the optical theorem \cite{J1975} ({\it i.e.}, $(1/\pi\hbar)\Imm\{\alpha(\omega)\}\,|\Eb^{\rm ext}(0,\omega)|^2=\sigma_{\rm ext}(\omega)I(\omega)/\hbar\omega$, where $\sigma_{\rm ext}(\omega)=(4\pi\omega/c)\Imm\{\alpha(\omega)\}$ is the extinction cross section and $I(\omega)=(c/4\pi^2)|\Eb^{\rm ext}(0,\omega)|^2$ is the light intensity per unit frequency), whereas the remaining term originates in electron-light interference. The probabilities given above are derived for isotropic dipolar particles, but a similar analysis leads to expressions corresponding to a particle characterized by a polarizability tensor $\alpha(\omega)\,\uu\otimes\uu$ ({\it i.e.}, linear induced polarization along a certain direction $\uu$), for which the partial probabilities are given by eqs\ \ref{CLdip} and \ref{allgammas} by substituting $\Eb^{\rm ext}$ and $\Eb^{\rm el}$ by $\uu\cdot\Eb^{\rm ext}$ and $\uu\cdot\Eb^{\rm el}$, respectively.

We explore the aforementioned energy pathways in Figure\ \ref{Fig5}b-e, where we plot the frequency-integrated probabilities $\Gamma_{\rm rad}$, $\Gamma_{\rm forward}$, $\Gamma_{\rm decay}$, and $\Gamma_{\rm el}$, respectively, as a function of electron-light pulse delay $\tau$. We find that the decay probability follows a similar symmetric profile as the radiative emission ({\it cf.} panels (b) and (d), both of them independent of the sign of $\tau$). In contrast, the electron energy-change probability (Figure\ \ref{Fig5}e) is markedly asymmetric (and so is the forward-emission probability (Figure\ \ref{Fig5}c) as a result of energy conservation via eq\ \ref{balance}): we find the intuitive result that the electron energy remains nearly unmodulated if the electron arrives before the optical pulse, while the opposite is true for the forward light emission component.

\section{Concluding Remarks}

Electron-beam-based spectroscopy techniques provide unrivalled spatial resolution for imaging sample excitations by measuring electron energy losses (EELS) or light emission (CL) associated with them. In this study, we propose the opposite approach: suppression of sample excitations produced by free electrons through combining them with mutually coherent laser irradiation. Indeed, our first-principles theory confirms that electrons and light can both be treated as mutually coherent tools for producing optical excitations. They form a synergetic team that combines optical spectral selectivity with the high spatial precision of electron beams. In contrast to EELS, where free electrons act as a broadband electromagnetic source, so that only {\it a posteriori} selection of specific mode frequencies is performed by spectrally resolving the inelastically scattered probes, the methods here explored allow us to target designated mode frequencies with sub-{\AA}ngstrom control over the excitation process. In addition, the excitation of on-demand nanoscale optical modes through the combined use of modulated electrons and tailored light pulses is amenable to the implementation of coherent control schemes \cite{paper080,BGM18} for the optimization of the desired effects on the specimen.

From a practical viewpoint, PINEM interaction provides a way of moulding the electron wave function to produce the temporally compressed pulses that are required to address specific sample frequencies. However, this method has a limited degree of achievable coherence in the electron-driven excitation process when using quasi-monochromatic light, quantified through the degree of coherence \cite{paperarxiv3} $0<{\rm DOC}(\omega)=|M_{\omega/v}|^2\le1$; more precisely, it can produce values ${\rm DOC}(\omega)\lesssim34\%$, as we show above. We remark that the frequency-dependent function ${\rm DOC}(\omega)$ is a property of the electron: this function is univocally determined by the probability density profile. Full coherence at a frequency $\omega$, corresponding to the ${\rm DOC}(\omega)\rightarrow1$ limit, can be delivered by $\delta$-function-like combs of electron pulses ({\it i.e.}, for an electron probability density $|\psi(z)|^2\approx\sum_m b_m\,\delta(z-2\pi mv/\omega)$ along the beam \cite{paperarxiv2}, with arbitrary coefficients $b_m$, including single pulses for $b_m=\delta_{m,0}$), the synthesis of which emerges as a challenge for future research.

By putting free electrons and light on a common basis as tools for creating excitations in a specimen, one could additionally envision the combined effect of multiple electron and laser pulses, which would increase the overall probability of exciting an optical mode, provided that their interactions take place within a small time interval compared with the mode lifetime. This idea capitalizes on the concept of superradiance produced by PINEM-modulated electrons \cite{GY20}, which our first-principles theory supports for probing and manipulating nanoscale excitations including the extra degrees of freedom brought by synchronized light and electron probes.

We remark that CL is just one instance of sample excitation, but the present study can be straightforwardly extended to optically bright modes in general (see independent analysis in ref\, \cite{paperarxiv2}), including two-level resonances of different multipolar character. A key ingredient of our work is the use of dimmed illumination, so that the weak probability amplitude that the electron typically imprints on the sample has a magnitude that is commensurate with the effect of the external light. Because the measurement is performed once interference between electron- and light-driven excitation amplitudes takes place ({\it i.e.}, at the far-field photospectrometer in CL, or by the effect of any subsequent inelastic process following the decay of the excited sample mode in general), the studied electron-light mutual coherence is unaffected by additional sources of shot noise other than the intrinsic ones associated with the detection process ({\it e.g.}, like in conventional CL).

Our prediction of unity-order effects in the modulation of electron-sample interactions through the use of external light enables applications in the manipulation of optical excitations at the atomic scale. Additionally, it suggests an alternative approach to damage-free sensing, whereby the spectral response of a specimen could be monitored through the modulation produced by the combined action of light and electrons, involving a reduced level of sample exposure to electrons because the targeted interference is proportional to the polarization amplitudes that they induce, so the outcome of a weak electron interaction could be amplified by applying a lock-in technique to the laser. This approach could be useful for imaging biomolecules, as well as strongly correlated materials in which probing without invasively perturbing the system is essential and remains a challenge in the exploration of spin and electronic ultrafast dynamics. Besides the experimental configuration proposed in Figure\ \ref{Fig1}, one could alternatively flip the semitransparent mirror horizontally to mix the external laser light with the CL emission at the detector instead of undergoing scattering at the specimen.

We find it interesting the possibility of adjusting the amplitude of the external light field, for example through a temporal light shaper, to determine the frequency-dependent magnitude and phase of the CL amplitude field ($\fb_{\rr}^{\rm CL}(\Rb,\omega)$ in our formalism), thus providing temporal resolution when probing the specimen by direct Fourier transformation of this quantity. This method could yield a time resolution limited by the width of the frequency window in the CL measurement at the spectrometer, without affecting the intrinsic temporal resolution associated with the short duration of electron and light pulses, and likewise, retaining the sub-{\AA}gnstrom spatial resolution associated with tightly focused electron beams. In a related direction, spatial light modulation and raster scanning of the electron beam could also be employed to gain further insight into the symmetry and nanoscale spatial dependence of the sample response. Additionally, for a sample in which $\fb_{\rr}^{\rm CL}(\Rb,\omega)$ is well characterized ({\it e.g.}, a dielectric sphere \cite{paperxx3}), the modulation of CL by varying the external field could be used to resolve the coherence factor $M_{\omega/v}$, thus allowing us to retrieve the electron density profile from the Fourier transform of this quantity. Besides far-field optical measurements, the present analysis can also be extended to alternative ways of probing optical excitations that are coherently created by light and electrons, such as electrical or acoustic detection of the modifications produced in the specimen.

\onecolumngrid
\section*{APPENDIX} 
\renewcommand{\thesection}{A} 
\renewcommand{\theequation}{A\arabic{equation}} 


\subsection{Quantization of the Electromagnetic Field in the Presence of Material Structures}

We follow ref\ \citenum{DKW98} for the quantization of the electromagnetic field in the presence of linearly responding materials characterized by a position- and frequency-dependent local permittivity $\epsilon(\rb,\omega)$. Without loss of generality to deal with free electrons that do not traverse any material, we adapt this formalism to a gauge in which the scalar potential is zero, as detailed elsewhere \cite{paper357}. The response of the media is represented through a noise current distribution operator $\hat{\jb}^{\rm noise}(\rb,\omega)$, in terms of which the vector potential operator reduces to
\begin{align}
\hat{\Ab}(\rb,\omega)=-4\pi c\int d^3\rb'\, G(\rb,\rb',\omega)\cdot\hat{\jb}^{\rm noise}(\rb',\omega),
\label{vpot}
\end{align}
where $G(\rb,\rb',\omega)$ is the classical electromagnetic Green tensor at frequency $\omega$, implicitly defined by eq\ \ref{greentensor}. The noise operator is chosen to be bosonic and satisfy the fluctuation-dissipation theorem for the current. These two conditions are fulfilled by writing
\begin{align}
\hat{\jb}^{\rm noise}(\rb,\omega)=\omega \sqrt{\hbar\,{\rm Im}\{\epsilon(\rb,\omega)\}}\;\hat{\fb}(\rb,\omega)
\label{cnoise}
\end{align}
in terms of bosonic ladder operators $\hat{\fb}(\rb,\omega)$ satifsying the commutation relations
\begin{subequations} 
\label{ffcom}
\begin{align}
[\hat{f}_i(\rb,\omega),\hat{f}_{i'}(\rb',\omega')]&=0,\\
[\hat{f}_i(\rb,\omega),\hat{f}_{i'}^\dagger (\rb',\omega')]&=\delta_{i,i'}\delta(\rb-\rb')\delta(\omega-\omega'),
\end{align}
\end{subequations}where $\hat{f}_{i=x,y,z}$ denotes the Cartesian components of $\hat{\fb}$. The Hamiltonian governing the free evolution of the radiation degrees of freedom is then expressed in terms of these operators as $\hat{\mathcal{H}}_{\rm rad}=\int d^3\rb \int_0^\infty d\omega \;\hbar\omega\; \hat{\fb}^\dagger (\rb,\omega)\cdot \hat{\fb}(\rb,\omega)$. By using eqs\ \ref{vpot} and \ref{cnoise}, the time-dependent quantum vector potential takes the form 
\begin{align}
\hat{\Ab}(\rb,t)= \int_0^\infty \frac{d\omega}{2\pi}\hat{\Ab}(\rb,\omega)\,\ee^{-\ii \omega t}+{\rm h.c.}
\label{vpott}
\end{align}
Of particular interest for the rest of the calculation are the different-time commutators between the quantum electromagnetic vector potential and the fields. These quantities can easily be obtained by using eqs\ \ref{vpot} to \ref{vpott}, together with the relations $\hat{\Eb}(\rb,t)=(-1/c)\,\partial_t \hat{\Ab}(\rb,t)$ and $\hat{\Bb}(\rb,t)=\nabla \times \hat\Ab(\rb,t)$, which lead to
\begin{subequations} 
\label{commutators}
\begin{align}
\left[\hat{B}_i(\rb,t),\hat{A}_{i'}(\rb',t')\right]&=8\ii c^2\hbar \int_0^\infty d\omega\;\sin[\omega(t-t')]\sum_{i''i'''} \epsilon_{ii''i'''}\; {\rm Im}\left\{\partial_{r_{i''}}G_{i'''i'}(\rb,\rb',\omega)\right\},\\
\left[\hat{E}_i(\rb,t),\hat{A}_{i'}(\rb',t')\right]&=-8\ii c\hbar \int_0^\infty \omega\, d\omega\;\cos[\omega(t-t')] {\rm Im}\left\{G_{ii'}(\rb,\rb',\omega)\right\}.
\end{align}
\end{subequations}
Here, we use the Levi-Civita symbol $\epsilon_{ii''i'''}$, as well as the identity \cite{DKW98}
\begin{align}
\sum_{i''}\int d^3\rb''\, {\rm Im}\{\epsilon(\rb'',\omega)\}\;G_{ii''}(\rb,\rb'',\omega)\,G^*_{i'i''}(\rb',\rb'',\omega)=-\frac{1}{\omega^2}{\rm Im}\left\{G_{ii'}(\rb,\rb',\omega)\right\}.
\label{imepsilon}
\end{align}
It is important to remark that the commutators between fields and potentials are c-numbers, only dependent on the time difference $t-t'$. In the calculation of the CL emission probability, we also need the retarded Green tensors constructed from the commutators in eqs\ \ref{commutators} as
\begin{subequations}
\label{retgreen}
\begin{align}
G_{{\rm BA},ii'}^{\rm R}(\rb,\rb',t-t')&=-\frac{\ii}{4\pi c^2 \hbar}\left[\hat{B}_i(\rb,t),\hat{A}_{i'}(\rb',t')\right]\theta(t-t'), \\
G_{{\rm EA},ii'}^{\rm R}(\rb,\rb',t-t')&=-\frac{1}{4\pi c \hbar}\left[\hat{E}_i(\rb,t),\hat{A}_{i'}(\rb',t')\right]\theta(t-t')
\end{align}
\end{subequations}
in the time domain, or equivalently,
\begin{subequations}
\label{retgreenfreq}
\begin{align}
G_{{\rm BA},ii'}^{\rm R}(\rb,\rb',\omega)&=\int_{-\infty}^\infty dt\,\ee^{\ii \omega t}\,G_{{\rm BA},ii'}^{\rm R}(\rb,\rb',t)=\sum_{i''i'''}\epsilon_{ii''i'''}\; \partial_{r_{i''}}G_{i'''i'}(\rb,\rb',\omega),\\
G_{{\rm EA},ii'}^{\rm R}(\rb,\rb',\omega)&=\int_{-\infty}^\infty dt\,\ee^{\ii \omega t}\,G_{{\rm EA},ii'}^{\rm R}(\rb,\rb',t)=\omega\,G_{ii'}(\rb,\rb',\omega)
\end{align}
\end{subequations}
in the frequency domain. In the derivation of eqs\ \ref{retgreenfreq}, we have used the fact that the electromagnetic Green tensor $G(\rb,\rb',\omega)$ satisfies the Kramers-Kronig relations and the causality property $G(\rb,\rb',-\omega)=G^*(\rb,\rb',\omega)$.

\subsection{Far-Field Radiation Emission: Derivation of Equation\ \ref{GEB}}

We now calculate the far-field emission produced by quantum currents taking into consideration the quantum nature of the electromagnetic excitations. To this aim, we define the average electromagnetic energy flow through a solid angular region $\Delta\Omega$ as 
\begin{align}
\Delta E = \lim_{kr\to\infty}\;r^2\int_{-\infty}^\infty dt \int_{\Delta\Omega} d^2\Omega_{\hat{\rb}}\, \big{\langle} \psi(-\infty)\big{|}\hat{\Sb}^{\rm H}(\rb,t) \cdot \hat{\rb}\big{|}\psi(-\infty)\big{\rangle}, \label{totenflow}
\end{align}
where $k=\omega/c$, $\hat{\Sb}^{\rm H}(\rb,t)=(c/8\pi)\,\left[\hat{\Eb}^{\rm H}(\rb,t)\times \hat{\Bb}^{\rm H}(\rb,t)- \hat{\Bb}^{\rm H}(\rb,t)\times \hat{\Eb}^{\rm H}(\rb,t)\right]$ is the quantum mechanical counterpart of the classical Poynting vector \cite{J1975}, and $|\psi(-\infty)\rangle$ is the initial quantum state at time $t=-\infty$. The superscript H indicates that operators have to be calculated in the Heisenberg picture, and thus evolved with the total Hamiltonian
\begin{align}
\hat{\mathcal{H}}_{\rm tot}=\hat{\mathcal{H}}_{\rm rad}+\hat{\mathcal{H}}_{\rm el}+\hat{\mathcal{H}}_{\rm int},
\nonumber
\end{align}
where $\hat{\mathcal{H}}_{\rm el}$ describes the free evolution of the electron degrees of freedom (or charge currents, in general) and $\hat{\mathcal{H}}_{\rm int}$ represents the light-currents interaction. Equation\ \ref{totenflow} can be expressed in terms of the scattering operator $\hat{\mathcal{S}}(t,-\infty)$ by incorporating an adiabatic switching of the interaction, which leads to the relation $\ee^{-\ii \hat{\mathcal{H}}_{\rm tot} t /\hbar}=\ee^{-\ii (\hat{\mathcal{H}}_{\rm rad}+\hat{\mathcal{H}}_{\rm el}) t/\hbar } \hat{\mathcal{S}}(t,-\infty) $ \cite{AGD1965}, and from here, eq\ \ref{totenflow} becomes
\begin{align}
\Delta E =\lim_{kr\to\infty}\;r^2 \int_{-\infty}^\infty dt \int_{\Delta\Omega} d^2\Omega_{\hat{\rb}}\, \langle \psi(-\infty) |\hat{\mathcal{S}}^\dagger (t,-\infty)\;\hat{\Sb}(\rb,t) \cdot \hat{\rb}\; \hat{\mathcal{S}}(t,-\infty) |\psi(-\infty)\rangle.
\label{totenflow2}
\end{align}
We now describe the interaction between the electromagnetic field and a total quantum current $\hat{\jb}(\rb,t)$ through the minimal coupling Hamiltonian in the zero scalar potential gauge as
\begin{align}
\hat{\mathcal{H}}_{\rm int}(t)=-\frac{1}{c}\int d^3\rb\; \hat{\Ab}(\rb,t)\cdot \hat{\jb}(\rb,t),
\label{hint}
\end{align}
where the time dependence in $\hat{\mathcal{H}}_{\rm int}(t)$ indicates that it is expressed in the interaction picture ({\it i.e.}, the free part of the Hamiltonian, $\hat{\mathcal{H}}_{\rm rad}+\hat{\mathcal{H}}_{\rm el}$, is taken care of through the scattering matrix). Because the commutator $\left[\hat{\Ab}(\rb,t),\hat{\Ab}(\rb',t')\right]$ is a c-number (this is a direct consequence of eqs\ \ref{vpot}-\ref{ffcom} and \ref{vpott}), if we assume that the current operators commute at different times and positions (see below), the scattering operator can be written as \cite{AGD1965,IZ12,paper357} (see detailed derivation below)
\begin{align}
\hat{\mathcal{S}}(t,-\infty)=\exp\left[\ii\,\hat{\chi}(t,-\infty)\right]\;\exp\left[-\frac{\ii}{\hbar}\int_{-\infty}^t dt'\, \hat{\mathcal{H}}_{\rm int}(t')\right],
\label{sop}
\end{align}
where the operator $\hat{\chi}(t,-\infty)$ only acts on the current degrees of freedom, and consequently, we can ignore it within the present work. From here, we plug eq\ \ref{sop} into eq\ \ref{totenflow2} and then use twice the identity $[\hat{A},\ee^{\hat{B}} ]=C\ee^{\hat{B}}$ (valid if $[\hat{A},\hat{B}]=C$ is a c-number) to bring the rightmost scattering operator to cancel its Hermitian conjugate on the left. This leads us to
\begin{align}
\Delta E = \lim_{kr\to\infty}\;&\frac{c\,r^2}{8\pi}\int_{-\infty}^\infty dt \int_{\Delta\Omega}
d^2\Omega_{\hat{\rb}}\; \Bigg \langle \left\{\hat{\Eb}(\rb,t)-\frac{\ii}{\hbar}\int_{-\infty}^tdt'\left[\hat{\Eb}(\rb,t),\hat{\mathcal{H}}_{\rm int}(t')\right]\right\} \nonumber\\
& \times \left\{\hat{\Bb}(\rb,t)-\frac{\ii}{\hbar}\int_{-\infty}^t dt'\left[\hat{\Bb}(\rb,t),\hat{\mathcal{H}}_{\rm int}(t')\right]\right\} \Bigg\rangle \cdot \rr+{\rm c.c.}, \label{totenflow3} 
\end{align}
where we have defined the quantum average as $\langle \cdot \rangle = \langle \psi(-\infty)|\cdot |\psi(-\infty) \rangle $. The term $\hat{\Eb}(\rb,t)\times\hat{\Bb}(\rb,t)$ in eq\ \ref{totenflow3}, which is independent of the sources, represents the contribution from the zero-point energy, so it bears no relevance to this analysis. In addition, since the commutators between the vector potential and the field operators are c-numbers, the terms linear in the currents ({\it i.e.}, through $\hat{\mathcal{H}}_{\rm int}$) in  eq\ \ref{totenflow3} vanish when they are averaged over an initial state $|\psi(-\infty)\rangle$ in which the radiation part is prepared in the photonic vacuum. Now, we use the retarded Green functions (eqs\ \ref{retgreen}) and their Fourier transforms (eqs\ \ref{retgreenfreq}) to obtain
\begin{align}
\Delta E = \int_0^\infty \hbar\omega\,d\omega \int_{\Delta\Omega} d^2\Omega_{\hat{\rb}} \; \frac{d\Gamma_{\rm ff}}{d\Omega_{\hat{\rb}}d\omega},
\nonumber
\end{align}
where
\begin{align}
\frac{d\Gamma_{\rm ff}}{d\Omega_{\hat{\rb}}d\omega}=\lim_{kr\to\infty}\;\frac{r^2}{4\pi^2\hbar k}{\rm Re}\left\{\left\langle\hat{\mathcal{E}}(\rb,\omega)\times \hat{\mathcal{B}}^\dagger(\rb,\omega) \right\rangle \right\}\cdot \hat{\rb}
\label{cldistrib}
\end{align}
is the angle- and frequency-resolved, time-integrated, far-field (ff) photon emission probability. Here, we have defined the new field operators
\begin{align}
\hat{\mathcal{E}}(\rb,\omega)=&-4\ii\pi\, \omega \int d^3\rb'\, G(\rb,\rb',\omega)\cdot\hat{\jb}(\rb',\omega), \nonumber\\
\hat{\mathcal{B}}(\rb,\omega)=&-4\pi\, c\; \nabla \times\int d^3\rb'\, G(\rb,\rb',\omega)\cdot\hat{\jb}(\rb',\omega), \nonumber
\end{align}
and we have introduced $\hat{\jb}(\rb,\omega)=\int_{-\infty}^{\infty} dt\,\ee^{\ii \omega t}\,\hat{\jb}(\rb,t) $. We note that eq\ \ref{cldistrib} resembles its classical counterpart \cite{paper149}, but now the currents are commuting quantum mechanical operators.

\subsection{Photon Intensity Produced by a Single Free Electron Combined with a Dimmed Laser: Derivation of Equation\ \ref{eqgen}}

We consider that the quantum current operator $\hat{\jb}$ is the sum of a classical term $\jb^{\rm ext}$ ({\it i.e.}, the source of the external laser light) and the quantum part associated with the free electrons $\hat{\jb}^{\rm el}$. For a highly energetic electron with central relativistic energy $E_0=c\sqrt{\me^2c^2+\hbar^2q_0^2}$ and initial wave function consisting of momentum components that are tightly focused around a central value $\hbar\qb_0$, the free-electron Hamiltonian $\hat{\mathcal{H}}_{\rm el}$ can be approximated as \cite{paper339} $\hat{\mathcal{H}}_{\rm el}=\sum_\qb\left[E_0+\hbar \vb \cdot (\qb-\qb_0)\right]\,\hat{c}_{\qb}^\dagger \hat{c}_{\qb}$, where $\vb=\hbar c^2\qb_0/E_0$ is the central electron velocity and we have introduced anticommuting creation and annihilation operators $\hat{c}_{\qb}^\dagger$ and $\hat{c}_{\qb}$ of an electron with momentum $\hbar\qb$. We remind that the momentum operator, written in the space basis set as $-\ii\hbar\nabla$ in ref\ \citenum{paper339}, now becomes $\sum_\qb \hbar\qb \hat{c}_{\qb}^\dagger \hat{c}_{\qb}$ in the second quantization formalism that we use here. Then, the electron current reduces to
\begin{align}
\hat{\jb}^{\rm el}(\rb,t)=-\frac{e \vb}{L^3}\sum_{\qb,\kb}\ee^{\ii \kb\cdot (\rb - \vb t)}\, \hat{c}_{\qb}^\dagger \hat{c}_{\qb+\kb},\label{qecurrent}
\end{align}
where $L$ is the side length of the quantization box, so wave vector sums can be transformed into integrals using the prescription $\sum_\qb\rightarrow(L/2\pi)^3\int d^3\qb$. By repeatedly using the anticommutation relations to pull all electron creation operators to the left, we find the commutation relation
\begin{align}
\left[\hat{\jb}^{\rm el}(\rb,t),\hat{\jb}^{\rm el}(\rb',t')\right]=0,
\label{commjj}
\end{align}
which is a property used above in the derivation of eq\ \ref{cldistrib}. Without loss of generality, we take $\vb$ along the $z$ and calculate the Fourier transform
\begin{align}
\hat{\jb}^{\rm el}(\rb,\omega)= -\zz\;\frac{e}{L^2}\ee^{\ii\omega z/v}\sum_{\qb,\kb_\perp}\ee^{\ii \kb_\perp\cdot\Rb}\,\hat{c}^\dagger_{\qb}\hat{c}_{\qb+\kb_\perp+(\omega/v)\zz},
\label{electroncurrent}
\end{align}
where $\kb_\perp\perp\zz$ is the transverse component of the exchanged wave vector $\kb$. This allows us to evaluate the average in eq\ \ref{cldistrib} for an initial state consisting of an electron prepared in a wave function $\psi^0(\rb)=\sum_{\qb}\alpha_{\qb}\langle\rb|\hat{c}_\qb^\dagger|0\rangle$ and zero photons ({\it i.e.}, $|\psi(-\infty)\rangle=\sum_{\qb}\alpha_{\qb}\hat{c}_\qb^\dagger|0\rangle$) by first computing the intermediate results
\begin{subequations}
\label{avcurrent}
\begin{align}
\langle \hat{\jb}^{\rm el}(\rb',\omega)\hat{\jb}^{{\rm el}\dagger }(\rb'',\omega) \rangle&=e^2\;\zz\otimes\zz\;\delta(\Rb'-\Rb'')\ee^{\ii \omega (z'-z'')/v}M_0(\Rb'),\\
\langle \hat{\jb}^{\rm el}(\rb',\omega) \rangle&=-e\;\zz\;\ee^{\ii\omega z'/v} M^*_{\omega/v}(\Rb'),
\label{avcurrent2}
\end{align}
\end{subequations}
where we use the notation $\rb=(\Rb,z)$. Also, $M_{\omega/v}(\Rb)$, defined in eq\ \ref{MwvR}, is a coherence factor that captures the dependence on the electron wave function through the probability density $|\psi^0(\rb)|^2$. We note that there is no dependence on the phase of $\psi^0(\rb)$. By using eqs\ \ref{avcurrent} to work out the evaluation of eq\ \ref{cldistrib}, we obtain
\begin{align}
\frac{d\Gamma_{\rm ff}}{d\Omega_{\hat{\rb}}d\omega}=\lim_{kr\to\infty}\;\frac{r^2}{4\pi^2\hbar k}{\rm Re}\Bigg\{&\frac{\ii}{k}\int d^2\Rb'\, M_0 (\Rb')\;\Eb^{\rm CL}(\rb,\Rb',\omega)\times \left[\nabla \times \Eb^{\rm CL}(\rb,\Rb',\omega)\right]^* \label{clff00}\\
& + \Eb^{\rm light}(\rb,\omega) \times \Bb^{{\rm light}*} (\rb ,\omega)\nonumber\\
&+\int d^2\Rb'\, M_{\omega/v}(\Rb')\; \Eb^{{\rm CL}*}(\rb,\Rb',\omega) \times \Bb^{\rm light}(\rb,\omega) \nonumber\\ 
&+\frac{\ii}{k}\int d^2\Rb'\, M_{\omega/v}(\Rb')\; \Eb^{\rm light}(\rb,\omega)\times \left[\nabla \times \Eb^{\rm CL}(\rb,\Rb',\omega)\right]^*\Bigg\}\cdot \hat{\rb},\nonumber
\end{align}
where we have defined the CL-related vector
\begin{align}
\Eb^{\rm CL}(\rb,\Rb',\omega)=4\pi\ii e\omega\int_{-\infty}^\infty dz'\, \ee^{\ii\omega z'/v}\,G(\rb,\Rb',z',\omega)\cdot \zz
\label{ECL}
\end{align}
and the total (external+scattered) light fields
\[\Eb^{{\rm light}}(\rb,\omega)=-4\pi\ii\, \omega \int d^3\rb'\; G(\rb,\rb',\omega)\cdot\jb^{\rm ext}(\rb',\omega)\]
and $\Bb^{{\rm light}}(\rb,\omega)=(-\ii/k)\,\nabla\times\Eb^{{\rm light}}(\rb,\omega)$. At this point, it is convenient to separate the light field into external and scattered components as $\Eb^{{\rm light}}(\rb,\omega)=\Eb^{{\rm ext}}(\rb,\omega)+\Eb^{{\rm scat}}(\rb,\omega)$, where the first term arises from the free-space part of the Green tensor, whereas the second-term decays as $1/r$ far from the sample. We consider first emission directions in which the external light does not interfere with the scattered and CL fields. Then, in the far-field limit ($kr\gg1$), we can approximate $\nabla\approx\ii k\rr$ in the above expressions, and the electric and magnetic fields only retain components perpendicular to $\rb$. This allows us to rewrite eq\ \ref{clff00} in the form given by eq\ \ref{eqgen} in terms of far-field electric field amplitudes $\fb_{\rr}^{\rm CL}(\Rb',\omega)$ and $\fb_{\rr}^{\rm scat}(\omega)$ associated with CL emission and laser scattering contributions (see definitions in eqs\ \ref{asympfs}). Under typical electron microscope conditions, for a well-focused electron beam, we can factorize the electron wave function as $\psi^0(\rb)=\psi_\perp(\Rb)\psi_\parallel(z)$ and approximate $|\psi_\perp(\Rb)|^2\approx\delta(\Rb-\Rb_0)$, where $\Rb_0$ defines the beam position. Inserting this wave function into eq\ \ref{eqgen}, we find
\begin{align}
\frac{d\Gamma_{\rm rad}(\Rb_0)}{d\Omega_{\hat{\rb}}d\omega}=\frac{1}{4\pi^2\hbar k}\bigg{[}|\fb_{\rr}^{\rm CL}(\Rb_0,\omega)|^2
+|\fb_{\rr}^{\rm scat}(\omega)|^2
+2\,{\rm Re}\left\{M_{\omega/v}\,\fb_{\rr}^{{\rm CL}*}(\Rb_0,\omega)\cdot\fb_{\rr}^{\rm scat}(\omega)\right\}\bigg{]},
\label{clfffocused}
\end{align}
where now $M_{\omega/v}$ is defined in eq\ \ref{Mwv}.

There is an additional component in $d\Gamma_{\rm ff}/d\Omega_{\hat{\rb}}d\omega$ (eq\ \ref{clff00}) arising from the interference between the external light field $\Eb^{\rm ext}(\rb,\omega)$ and the scattered+CL far-field amplitudes. For plane wave light incidence with wave vector $\kb_{\rm inc}$, the former can be written $\Eb^{\rm ext}(0,\omega)\ee^{\ii\kb_{\rm inc}\cdot\rb}$, which contributes to $d\Gamma_{\rm ff}/d\Omega_{\hat{\rb}}d\omega$ through the three last terms of eq\ \ref{clff00}. After integration over emission directions, and considering a dipolar scatterer (see below), this contribution becomes $d\Gamma_{\rm forward}/d\omega$ (eq\ \ref{forward}) (see Supplementary Information for more details).

\subsection{Generalization to Multiple Electrons: Derivation of Equation\ \ref{multielectrons}}

The above formalism can be readily extended to deal with more than one electron by taking the initial state as $|\psi(-\infty)\rangle=\prod_j \left(\sum_{\qb_j}\alpha^j_{\qb_j}c^\dagger_{\qb_j}\right)|0\rangle$, where $j$ runs over different electrons and the photonic field is prepared in the vacuum state. Then, using the definition of the electron current operator $\hat{\jb}^{\rm el}(\rb,\omega)$ in eq\ \ref{electroncurrent}, the averages in eqs\ \ref{avcurrent} can be readily computed for the multi-electron state to yield
\begin{subequations}
\label{avcurrentmulti}
\begin{align}
\langle\hat{\jb}^{\rm el}(\rb',\omega)\hat{\jb}^{\rm el\dagger }(\rb'',\omega)\rangle &=e^2\,\zz\otimes\zz\, \ee^{\ii \omega (z'-z'')/v} \nonumber\\
&\times\left[  \delta(\Rb'-\Rb'') \sum_j M^j_0(\Rb')+\sum_{j\neq j'} M^{j*}_{\omega/v}(\Rb')M^{j'}_{\omega/v}(\Rb'')\right],\\
\langle\hat{\jb}^{\rm el}(\rb',\omega)\rangle&=-e\,\zz\, \ee^{\ii \omega z'/v}\sum_j M^{j*}_{\omega/v}(\Rb'),
\end{align}
\end{subequations}
where $M^j_{\omega/v}$ is given by eq\ \ref{Mwv} with $\psi^0(\rb)$ substituted by $\psi^j(\rb)=\sum_{\qb}\alpha^j_{\qb}\langle\rb|\hat{c}_\qb^\dagger|0\rangle$ (the wave function of electron $j$). Finally, plugging eqs\ \ref{avcurrentmulti} into eq\ \ref{cldistrib} and following similar steps as done above for a single electron, we obtain eq\ \ref{multielectrons} in the main text.

\subsection{Cathodoluminescence from a Dipolar Sample Object: Derivation of Equation\ \ref{CLdip}}

We present results in the main text for sample objects whose response are dominated by an electric dipolar mode represented through an isotropic polarizability $\alpha(\omega)$ placed at $\rb=0$. We now carry out the limit in eq\ \ref{limitvectorq} by realizing that the free-space component of the Green tensor to the $z'$ integral vanishes exponentially away from the electron beam ({\it i.e.}, just like the electromagnetic field accompanying a freely moving classical charge), so we only need to account for the contribution from the scattering part,
\begin{align}
G^{\rm scat}(\rb,\rb',\omega)\xrightarrow[kr\to\infty]{} -\frac{\alpha(\omega)}{4\pi c^2}\frac{\ee^{\ii kr}}{r}(1- \hat{\rb }\otimes \hat{\rb})\cdot(k^2+\nabla_{\rb '} \otimes \nabla_{\rb '})\frac{\ee^{\ii kr'}}{r'}.
\nonumber
\end{align}
Plugging this expression into eq\ \ref{limitvectorq}, we can carry out the $z'$ integral by using the identities $\int_{-\infty}^\infty dz\, \ee^{\ii\omega(z/v+r/c)}/r=2K_0\left(\omega R/v\gamma\right)$ and $\int_{-\infty}^\infty dz\,(1+\ii/kr)\,\ee^{\ii\omega(z/v+r/c)}/r^2=(2\ii c/R v\gamma)\,K_1\left(\omega R/v\gamma\right)$, where $r=\sqrt{R^2+z^2}$ and $\gamma=1/\sqrt{1-v^2/c^2}$ (see eqs\ 3.914-4 and 3.914-5 in ref\ \citenum{GR1980}). This leads to
\begin{align}
\fb_{\rr}^{\rm CL}(\Rb',\omega)=k^2\,\alpha(\omega)\,\left(1-\rr\otimes\rr\right)\cdot\Eb^{\rm el}(\Rb',\omega),
\label{scattelfield}
\end{align}
where $\Eb^{\rm el}(\Rb',\omega)$, defined in eq\ \ref{FF}, coincides with the electric field produced at the particle position $\rb=0$ by a classical point electron whose trajectory crosses $(\Rb',0)$ at time $t=0$ \cite{paper149}. Similarly, from eq\ \ref{limitvectorf}, the scattered external field amplitude is readily found to be
\begin{align}
\fb_{\rr}^{\rm scat}(\omega)=k^2\,\alpha(\omega)\,\left(1-\rr\otimes\rr\right)\cdot\Eb^{\rm ext}(0,\omega),\label{sextfield}
\end{align}  
where $\Eb^{\rm ext}(0,\omega)$ is the external laser field acting on the particle. Finally, by inserting eqs\ \ref{scattelfield} and \ref{sextfield} into eq\ \ref{clfffocused}, we obtain
\begin{align}
\frac{d\Gamma_{\rm rad}}{d\Omega_{\hat{\rb}}d\omega}&=\frac{k^3}{4\pi^2\hbar}|\alpha(\omega)|^2\Big\{\left(|\Eb^{\rm el}(\Rb_0,\omega)|^2-|\rr\cdot\Eb^{\rm el}(\Rb_0,\omega)|^2\right) + \left(|\Eb^{\rm ext}(0,\omega)|^2-|\rr\cdot\Eb^{\rm ext}(0,\omega)|^2\right) \nonumber \\
&+2\, {\rm Re}\left\{M_{\omega/v} \left[\Eb^{{\rm el}*}(\Rb_0,\omega)\cdot\Eb^{\rm ext}(0,\omega)-\left(\rr\cdot\Eb^{{\rm el}*}(\Rb_0,\omega)\right) \left(\rr\cdot\Eb^{\rm ext}(0,\omega)\right) \right] \right\}\Big\}.
\label{angularcldipfoc}
\end{align}
The total far-field photon probability per unit frequency is then obtained by integrating eq\ \ref{angularcldipfoc} over solid angles, leading to
\begin{align}
\frac{d\Gamma_{\rm rad}(\Rb_0)}{d\omega}=& \frac{2k^3}{3\pi\hbar} |\alpha(\omega)|^2  \nonumber\\
&\times \left[ |\Eb^{\rm el}(\Rb_0,\omega)|^2 + |\Eb^{\rm ext}(0,\omega)|^2+2 {\rm Re}\left\{M_{\omega/v}\Eb^{{\rm el}*}(\Rb_0,\omega)\cdot \Eb^{\rm ext}(0,\omega)\right\}\right].
\label{dipclf}
\end{align}
This expression can readily be recast in the form of eq\ \ref{CLdip} in the main text.

\subsection{The Scattering Operator: Derivation of Equation\ \ref{sop}}

We describe our system through the interaction Hamiltonian in eq\ \ref{hint} and use the commutation relation in eq\ \ref{commjj} to write $\left[\hat{\mathcal{H}}_{\rm int}(t),\hat{\mathcal{H}}_{\rm int}(t')\right]=(1/c^2)\int d^3 \rb d^3 \rb '\, \hat{\jb}^{\rm e} (\rb,t)\cdot \left[\hat{\Ab}(\rb,t),\hat{\Ab} (\rb',t')\right]\cdot \hat{\jb}^{\rm e}(\rb',t')$. Additionally, eqs\ \ref{vpot} and \ref{ffcom} directly imply that $\left[\hat{\Ab} (\rb,t),\hat{\Ab} (\rb',t')\right]$ is a c-number, which in turn leads to the nested commutation relation
\begin{align}
\left[\hat{\mathcal{H}}_{\rm int}(t''),\left[\hat{\mathcal{H}}_{\rm int}(t),\hat{\mathcal{H}}_{\rm int}(t')\right]\right]=0.\label{ncommr}
\end{align}
This expression is important to derive eq\ \ref{sop} for the scattering operator starting from its definition \cite{AGD1965} $\hat{\mathcal{S}}(t,t_0)=T\exp\left[(-\ii/\hbar)\int_{t_0}^t dt'~\hat{\mathcal{H}}_{\rm int}(t')\right]$, where $T$ denotes time ordering. Following a well-established procedure \cite{IZ12}, we discretize the time integral (with a set of equally spaced times $t_i$ with $i=1,\dots,N$) and explicitly implement time ordering to write
\begin{align}
\hat{\mathcal{S}}(t,t_0)&=\lim_{N\rightarrow \infty} \ee^{(-\ii/\hbar)\Delta t \hat{\mathcal{H}}_{\rm int}(t_N)} \ee^{(-\ii/\hbar)\Delta t \hat{\mathcal{H}}_{\rm int}(t_{N-1})} \dots  \ee^{(-\ii/\hbar)\Delta t \hat{\mathcal{H}}_{\rm int}(t_{1})} \nonumber \\
&=\lim_{N\rightarrow \infty}\exp\bigg\{\frac{-\ii}{\hbar}\Delta t \sum_{i=1}^N  \hat{\mathcal{H}}_{\rm int}(t_{i})-\frac{\Delta t^2}{2\hbar^2}\sum_{1\leq k < l \leq N}\left[ \hat{\mathcal{H}}_{\rm int}(t_{l}), \hat{\mathcal{H}}_{\rm int}(t_{k})\right]\bigg\} \nonumber
\end{align}
with $\Delta t=(t-t_0)/N$, where we have used the relation $\ee^{\hat{X}}\ee^{\hat{Y}}=\ee^{\hat{X}+\hat{Y}+[\hat{X},\hat{Y}]/2}$, which is valid if $[\hat{X},[\hat{X},\hat{Y}]]=[\hat{Y},[\hat{X},\hat{Y}]]=0$ ({\it i.e.}, like in eq\ \ref{ncommr}). Using this identity again, we readily find eq\ \ref{sop} by setting $t_0=-\infty$ and defining the phase operator $\hat{\chi}(t,-\infty)=(\ii/2 \hbar^2)\int_{-\infty}^t dt' \int_{-\infty}^t dt''\,\theta(t'-t'')\left[\hat{\mathcal{H}}_{\rm int}(t'),\hat{\mathcal{H}}_{\rm int}(t'')\right]$. Interestingly, since the commutator between the electromagnetic potentials is a c-number, the operator $\hat{\chi}(t,-\infty)$ acts only on the degrees of freedom associated with the currents and represents the effect of the image potential acting on the free charges \cite{paper357}.

One is often interested in calculating asymptotic quantities such as electron spectra at $t=\infty$. We then need to know the scattering operator $\hat{\mathcal{S}}(\infty,-\infty)$, which can be obtained by using eqs\ \ref{vpot} and \ref{qecurrent}, leading to $\hat{\mathcal{S}}(\infty,-\infty)=\exp\left[\ii \hat{\chi}(\infty,-\infty)\right]\hat{\mathcal{U}}$, where
\begin{align}
\hat{\mathcal{U}}=\exp\bigg\{\bigg[\frac{-\ii e}{2\pi\hbar c L^2}\sum_{\qb ,\kb_{\perp}}\int_0^\infty  d\omega \int d^3\rb\,\ee^{\ii \kb_\perp \cdot \Rb} \ee^{-\ii \omega z/v } \hat{A}_z(\rb,\omega)\, \hat{c}^\dagger_{\qb}\hat{c}_{\qb+\kb_{\perp}-(\omega/v)\zz}\bigg]-{\rm h.c.}\bigg\} \nonumber
\end{align}
(see definition of $\hat\Ab$ in eq\ \ref{vpot}) describes the total time evolution of electron-light states in the nonrecoil approximation if we disregard the effect of the image potential ({\it i.e.}, the phase operator $\hat{\chi}$). When the electron is focused around a point $\Rb=\Rb_0$ and its wave function can be separated in longitudinal and transverse components, as we do in the main text, we can approximate $\hat{c}_\qb \approx \hat{c}_{\qb_\perp}\hat{c}_{q_z}$ and replace the operator in the exponent of $\hat{\mathcal{U}}$ by its average over a transverse electron state $|\psi_\perp\rangle= \sum_{\qb_\perp}\alpha_{\qb_\perp}|\qb_\perp\rangle$ satisfying the relation $\sum_{\kb_\perp}\alpha_{\kb_\perp} \alpha^*_{\kb_\perp+\qb_\perp} = \ee^{\ii \qb_\perp \cdot \Rb_0}$, from which we find
\begin{align}
\hat{\mathcal{U}}=\exp\left[\int_0^\infty d\omega\, g_\omega (\hat{b}_\omega^\dagger \hat{a}_\omega-\hat{b}_\omega \hat{a}^\dagger_\omega )\right]. \label{uscaled}
\end{align} 
Here, we have introduced the operators $\hat{a}_\omega=(-\ii e/2\pi\hbar c\,g_\omega) \int_{-\infty}^\infty dz \ee^{- \ii \omega z /v }\hat{A}_z(\Rb_0,z,\omega)$ and $\hat{b}_\omega=\sum_{q_z}\hat{c}^\dagger_{q_z} \hat{c}_{q_z+\omega/v}$, as well as the coupling coefficient $g_\omega=\sqrt{\Gamma_{\rm EELS}(\Rb_0,\omega)}$, which reduces to the square root of the classical EELS probability \cite{paper149} \[\Gamma_{\rm EELS}(\Rb_0,\omega)=(4 e^2/\hbar)\int_{-\infty}^\infty dz \int_{-\infty}^\infty dz' \cos\left[\omega(z-z^\prime)/v\right] {\rm Im}\left\{-G_{zz}(\Rb_0,z,\Rb_0,z',\omega)\right\}.\] We define these operators in such a way that they satisfy the commutation relations $[\hat{a}_\omega,\hat{a}^\dagger_{\omega^\prime}]=\delta(\omega-\omega^\prime)$ and $[\hat{b}_\omega,\hat{b}^\dagger_{\omega^\prime} ]=0$, where the former can be proven by using eq\ \ref{imepsilon}. Importantly, eq\ \ref{uscaled} allows us to quickly compute observables after electron-sample interaction. As an example of this, we find that the average of the positive-energy electric field operator $\hat{\Eb}^{(+)}(\rb,\omega)=\ii k\hat{\Ab}(\rb,\omega )$ over the state  $|\psi(\infty)\rangle=\hat{\mathcal{S}}(\infty,-\infty)|\psi(-\infty)\rangle$ with $|\psi(-\infty)\rangle=\sum_{q_z} \alpha_{q_z}\hat{c}_{q_z}^\dagger|0\rangle$ (proportional to the photonic vacuum) reduces to $\langle \hat{\Eb}^{(+)}(\rb,\omega) \rangle =8\pi e \omega\,\Gb(\rb, \omega) M_{\omega/v}^*$, where $\Gb(\rb,\omega)=\int_{-\infty}^\infty dz'\, \ee^{\ii\omega z'/v} {\rm Im}\{G(\rb,\Rb_0,z',\omega)\}\cdot \zz $. To derive this result, we need to use the relation $[\hat{A},\ee^{\hat{B}} ]=C\ee^{\hat{B}}$ (valid if $[\hat{A},\hat{B}]=C$ is a c-number), as well as the commutation relation $[\hat{\fb}(\rb,\omega),\int_0^\infty d\omega ' g_{\omega'}(\hat{b}_{\omega'} \hat{a}_{\omega'}^\dagger-\hat{b}_{\omega'}^\dagger  \hat{a}_{\omega'})]=(-2 \ii e \omega/\hbar)\,\hat{b}_\omega\,\sqrt{\hbar {\rm Im}\{\epsilon (\rb ,\omega )\}}\int_{-\infty}^\infty dz'\,\ee^{\ii\omega z'/v}G(\rb,\Rb_0,z',\omega)\cdot \zz$ together with the fact that the fermionic operators $\hat{b}_\omega$ and $\hat{b}^\dagger_\omega$ commute.

\subsection{Calculation of the Coherence Factor for PINEM-Modulated Electrons}

For an electron whose wave function is the product of eqs\ \ref{Gaussian} and \ref{PPINEM}, the coherence factor defined in eq\ \ref{Mwv} readily reduces to the expression
\begin{align}
M_{\omega/v}=
\sum_{ll'} \ee^{-\sigma_t^2[(l-l')\omega_P+\omega]^2/2}J_l(2|\beta|)J_{l'}(2|\beta|)\,\ee^{\ii (l'-l)\omega_Pz_P/v+2\pi\ii(l^{\prime 2}-l^2)d/z_T},
\nonumber
\end{align}
which we evaluate numerically for finite $\sigma_t$. In the $\omega_P\sigma_t\gg1$ limit, $M_{\omega/v}$ takes negligible values unless the excitation frequency is a multiple of the PINEM laser frequency ({\it i.e.}, $\omega=m\omega_P$). Then, only $l'=l+m$ terms contribute to the above sum, which reduces to $M_{\omega/v}=\ee^{\ii m\omega_Pz_P/v+2\pi\ii m^2d/z_T}\sum_{l} J_l(2|\beta|)J_{l+m}(2|\beta|)\,\ee^{4\pi\ii mld/z_T}$ and using Graf's addition theorem, we have $|M_{\omega/v}|=|J_m[4|\beta|\sin(2\pi m d /z_T)]|$, in agreement with ref \cite{ZSF21}. We use this equation with $m=1$ to obtain the map shown in Figure\ \ref{Fig4}a, and with $m=1-3$ to produce the supplementary Figure\ S2.




\section*{}
\twocolumngrid
\section*{Acknowledgments} 

This work has been supported in part by the European Research Council (Advanced Grant 789104-eNANO), the European Commission (Horizon 2020 Grant 101017720 FET-Proactive EBEAM), the Spanish MINECO (MAT2017-88492-R and Severo Ochoa CEX2019-000910-S), the Catalan CERCA Program, the Fundaci\'{o}s Cellex and Mir-Puig, and the Humboldt Foundation. C.R. gratefully acknowledges funding by the Deutsche Forschungsgemeinschaft (DFG, German Research Foundation) from the Gottfried Wilhelm Leibniz prize (RO 3936/4-1) and {\it via} Priority Program 1840 'Quantum Dynamics in Tailored Intense Fields' (project No. 281311214). V.D.G. acknowledges support from the EU (Marie Sk\l{}odowska-Curie Grant 713729). O.K. acknowledges the Max Planck Society for a Manfred Eigen Fellowship for postdoctoral fellows from abroad.


\clearpage 
\pagebreak \onecolumngrid \section*{SUPPLEMENTARY INFORMATION} 
\renewcommand{\thefigure}{S\arabic{figure}} 
\renewcommand{\theequation}{S\arabic{equation}} 
\renewcommand{\thetable}{S\arabic{table}} 
\renewcommand{\thesection}{S\arabic{section}} 
\renewcommand{\thepage}{S\arabic{page}} 

\onecolumngrid
\section*{Coherence in Electron Microscopy}
\twocolumngrid

The term {\it coherence} is employed to denote different things depending on the physical processes under consideration. We thus provide a brief discussion on several possible uses of this concept in the context of electron microscopy.

{\bf Coherence in the sampled excitations.} Plasmons and other types of polaritons excited by electron beams (e-beams) can be out-coupled to radiation at different regions of the specimen, and eventually produce far-field CL interference. In this sense, the coherence of these excitations is a property of the specimen, independent of whether we excite them with fast electrons or with localized point emitters ({\it e.g.}, quantum dots). Even Smith-Purcell radiation fits into this general description: there is a continuum of degenerate light modes for each emission frequency ({\it i.e.}, different directions of emission and two polarizations) and the electron just provides a practical way of accessing a particular superposition of them ({\it i.e.}, the electron velocity and beam orientation relative to the grating select which modes are excited, corresponding to the emission of radiation along specific frequency-dependent directions).

Coherence in this context then refers to the interference between different excited modes (polaritons and photons) when the quantum mechanical state of the specimen remains unchanged after interaction with the electron \cite{paper149}. An example of incoherent excitations according to this definition is provided by those associated with the luminescence resulting from the decay of interband electronic transitions in a semiconductor that are initially created by the e-beam and subsequently undergo de-excitation to an intermediate state that introduces a random phase ({\it e.g.}, {\it via} an Auger process): the sample is not left in the same quantum-mechanical state after interaction with the electron, so the emission intensity builds up from the (incoherent) sum of intensities associated with different excitations triggered by the electron at different locations.

\begin{figure}[H]
\centering{\includegraphics[width=0.50\textwidth]{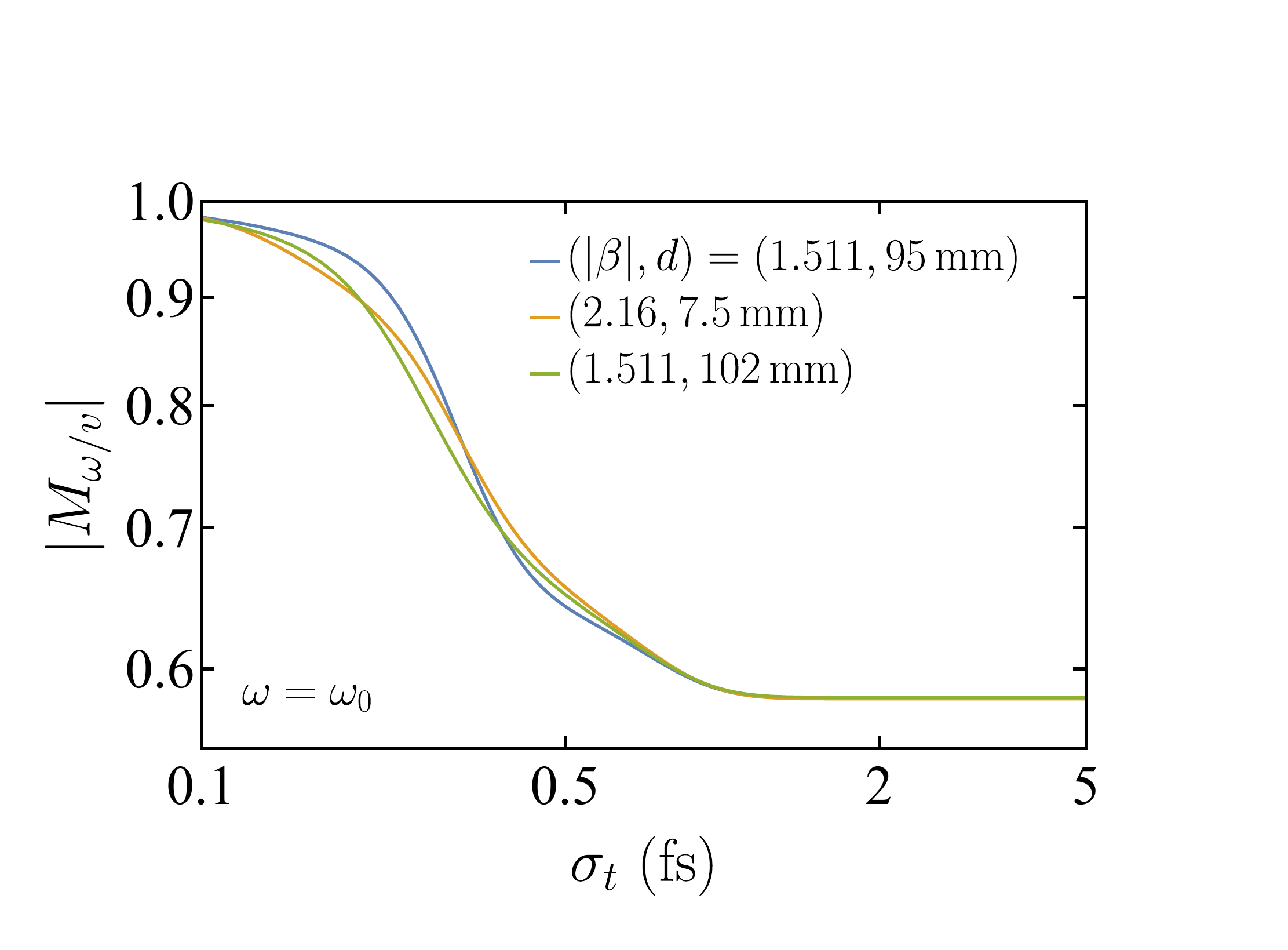}}
\caption{{\bf Dependence of the coherence factor on Gaussian wavepacket duration.} We show $|M_{\omega/v}|$ for a PINEM-modulated electron with three different combinations of parameters $|\beta|$ and $d$ (see legend) along the blue line of maxima in Figure\ 3a in the main text as a function of the duration $\sigma_t$ of a superimposed Gaussian wavepacket envelope. The excitation energy is $\hbar\omega_0=1.3\,$eV.}
\label{FigS1}
\end{figure}

{\bf Coherence of the electron as a source of optical excitations.} Coherence in this sense depends on the electron wave function (or in general the electron density matrix if it is prepared in a mixed state). It manifests during the interaction of the sample with several synchronized electrons (see point {\it iii} below) and also by means of interference of CL with external light (points {\it iv-v}). To complement this discussion, several additional elements related to coherence can be demonstrated from first principles \cite{paperarxiv2} ({\it i.e.}, with independence of the type of specimen and the mechanism of interaction with the electron) under the assumption of nonrecoil:
\begin{enumerate}[{\it i}.]
\item For interaction with an individual electron, the excitation probability for both EELS and CL reduces to the average of the probability $P(x,y)$ obtained for a classical point particle passing by $(x,y)$ over the e-beam transverse density profile \cite{RH1988} ({\it i.e.}, the resulting probability is $\iint dxdy \;|\psi_\perp(x,y)\,|^2P(x,y)$, where $\psi_\perp(x,y)$ is the lateral component of the wave function and the electron velocity is taken along $z$). This result was first obtained by Ritchie and Howie \cite{RH1988}.
\end{enumerate}
\begin{enumerate}[{\it ii}.]
\item The individual-electron probability $P(x,y)$ is actually independent of the longitudinal wave function $\psi_\parallel(z)$.
\end{enumerate}
\begin{enumerate}[{\it iii}.]
\item For interaction with more than one electron during the lifetime of the sampled excitation, the probability can depend on the electron wave functions if they are mutually synchronized. This dependence comes through factors given by $M_{\omega/v}$ (eq\ 9 in the main text).
\end{enumerate}
Additionally, we might wonder whether the excitations produced by the electron maintain some degree of coherence with respect to any external illumination. In the main text, we show that, in the limit of a point-particle electron, the CL emission is completely coherent with respect to external light if this is synchronized to a high precision relative to the optical cycle of the sampled excitation: the electron acts as a classical source that is phase-locked to the external light, so the generated far-field amplitude is the sum of contributions coming from each of them ({i.e.}, the solution of Maxwell's equations for the combined electron and light sources). Now, the question arises, is this also true when the electron is not a point particle? We find the following answers from first-principles theory:
\begin{enumerate}[{\it iv}.]
\item The CL emission can partially interfere with external light if the electron and the light are mutually phase-locked. Interference comes through a term proportional to the so-called {\it degree of coherence} $|M_{\omega/v}|^2$ \cite{paperarxiv3}, where $M_{\omega/v}$ is the same quantity that rules the interference between the excitations produced by multiple synchronized electrons \cite{paperarxiv2}). Maximum coherence corresponds to $|M_{\omega/v}|^2=1$, which can be achieved if the electron probability density consists of a series of $\delta$-function peaks separated by a distance $2\pi v/\omega$ \cite{paperarxiv2}.
\end{enumerate}
\begin{enumerate}[{\it v}.]
\item As we show in the main text, the CL emission can even be partially suppressed if it is mixed with mutually coherent light. The maximum fraction of emission that can be suppressed is given by $|M_{\omega/v}|^2$.
\end{enumerate}
\noindent For electrons prepared in Gaussian wavepackets, the factor $|M_{\omega/v}|^2=\ee^{-\omega^2\sigma_t^2}$ depends on their duration $\sigma_t$ relative to the optical cycle of the sampled excitation $2\pi/\omega$ ({\it e.g.}, $2\pi/\omega\approx4.1\,$fs for $\hbar\omega=1\,$eV). A practical way to achieve a significant value of $|M_{\omega/v}|^2$ consists in using PINEM-modulated electrons, and then the PINEM laser is automatically synchronized with the electron modulation. However, under cw illumination, this leads to $|M_{\omega/v}|^2<0.34$, unless the electron is already prepared as a short pulse (Figure\ \ref{FigS1}). The quest for achieving the maximum possible value of $|M_{\omega/v}|^2=1$ defines an exciting avenue of research.


\onecolumngrid
\section*{Alternative Description for a Dipolar Scatterer: Analysis of Energy Pathways}
\def\ww{\omega}  \def\w0{\omega_0}
\twocolumngrid

We present an alternative treatment of a dipolar scatterer that hosts a single optical mode. This approach does not require photon quantization and it can be applied to any two-level system that can be characterized by a transition dipole. As a starting point, we write the Hamiltonian
\begin{align}
\hat{\mathcal{H}}= &\hbar\w0\,\hat{a}^\dagger \hat{a} + \hbar\sum_q\varepsilon_q \hat{c}_q^\dagger \hat{c}_q
\label{H}\\&+g(t)\left(\hat{a}^\dagger+\hat{a}\right)
+\sum_{qq'}g_{qq'}\hat{c}_q^\dagger \hat{c}_{q'}\left(\hat{a}^\dagger+\hat{a}\right), \nonumber
\end{align}
where $\w0$ is the mode frequency, $\hat{a}^\dagger$ and $\hat{a}$ represent the corresponding creation and annihilation operators, $\hat{c}_q^\dagger$ and $\hat{c}_q$ create and annihilate an electron of wave vector $q$ and kinetic energy $\hbar\varepsilon_q$ along the e-beam direction, the real coefficient $g(t)$ describes the mode coupling to classical external light, and $g_{qq'}$ are electron-scatterer coupling coefficients.

In what follows, we ignore transverse coordinates under the nonrecoil approximation, together with the assumption that the e-beam is focused around a lateral position $\Rb_0=(x_0,y_0)$ relative to the scatterer, with a small focal spot compared to both $c/\w0$ and $R_0$. A basis set of longitudinal wave vector states $\langle z|q\rangle=\ee^{\ii qz}/\sqrt{L}$ is then used to describe the electron, where $L$ is the quantization length along the e-beam direction. In addition, the scatterer is considered to be prepared in its ground state before interaction with the external light and the electron. We further assume typical conditions in electron microscopy, characterized by a weak electron-scatterer interaction, so that we can work to the lowest possible order of perturbation theory. The external light is taken to be dimmed, such that its interaction strength becomes commensurate with that of the electron. Under these conditions, the density matrix of the combined electron-scatterer system can be written as
\begin{align}
\hat\rho=\sum_{nn',qq'}\alpha_{nn',qq'}(t)\;\ee^{\ii(n'-n)\omega_0t+\ii\varepsilon_{q'q}t}\;|nq\rangle\langle n'q'|,
\label{rho}
\end{align}
where $|nq\rangle\equiv(\hat{a}^\dagger)^n\hat{c}_q^\dagger|0\rangle/\sqrt{n!}$ and we adopt the notation $\varepsilon_{q'q}=\varepsilon_{q'}-\varepsilon_q$. A finite lifetime $\tau_0$ of the optical mode is now introduced through the equation of motion
\begin{align}
\frac{d\hat\rho}{dt}=\frac{\ii}{\hbar}\left[\hat\rho,\hat{\mathcal{H}}\right]
+\frac{1}{2\tau_0}\left(2\hat{a}\hat\rho \hat{a}^\dagger-\hat{a}^\dagger \hat{a}\hat\rho-\hat\rho \hat{a}^\dagger \hat{a}\right).
\label{rhot}
\end{align}
Before interaction, the coefficients of the density matrix are $\alpha_{nn',qq'}(-\infty)=\delta_{n0}\delta_{n'0}\alpha_q^0\alpha_{q'}^{0*}$, where $\alpha_q^0$ defines the incident longitudinal electron wave function
\begin{align}
\psi_\parallel(z)=\sum_q\alpha_q^0\langle z|q\rangle=\sqrt{L}\int_{-\infty}^\infty \frac{dq}{2\pi}\,\alpha_q^0\,\ee^{\ii qz}.
\label{psiz}
\end{align}
Here, we have used the prescription $\sum_q\rightarrow(L/2\pi)\int_{-\infty}^\infty dq$ to transform the sum over the electron wave vector $q$ into an integral.

We consider external light characterized by an electric field $\Eb^{\rm ext}(\rb,t)$ at the position of the scatterer, so we have
\begin{align}
g(t)=-\pb_0\cdot\Eb^{\rm ext}(0,t),
\label{gt}
\end{align}
where $\pb_0$ is the transition dipole. Additionally, the electron-scatterer coupling coefficients are given by \cite{paper221}
\begin{align}
g_{qq'}=g_{q'q}^*=-\frac{v}{L}\,\pb_0\cdot\gb_{q'-q},
\label{gqq}
\end{align}
where
\begin{align}
\gb_q=\frac{2e}{v\gamma}\bigg[&|q|\,K_1\left(|q|R_0/\gamma\right)\,\hat{\Rb}_0+\frac{\ii q}{\gamma}\,K_0\left(|q|R_0/\gamma\right)\,\zz\bigg], \nonumber
\end{align}
$v$ is the average electron velocity, and $\gamma=1/\sqrt{1-v^2/c^2}$.

\onecolumngrid

The excitation probabilities here investigated are determined by the diagonal elements $\alpha_{nn,qq}(t)$, which we calculate to the lowest order of perturbation theory by plugging eqs\ \ref{H} and \ref{rho} into eq\ \ref{rhot}. Identifying the coefficient of each $|nq\rangle\langle n'q'|$ term in both sides of the resulting equation, iteratively evaluating the correction to $\alpha_{nn',qq'}$ at perturbation order $l+1$ by inserting the order-$l$ correction into the $[\hat\rho,\hat{\mathcal{H}}]$ term of eq\ \ref{rhot}, and starting with $\alpha_{nn',qq'}(-\infty)$ for $l=0$ (see above), we find
\begin{subequations}
\begin{align}
&\frac{d\alpha_{01,qq'}(t)}{dt}=\frac{\ii}{\hbar}g(t)\,\alpha_q^0\,\alpha_{q'}^{0*}\;\ee^{-\ii\w0 t}
+\frac{\ii}{\hbar}\sum_{q''}g_{q''q'}\,\alpha_{q}^0\,\alpha_{q''}^{0*}\;\ee^{-\ii(\w0+\varepsilon_{q'q''})t}
-\frac{1}{2\tau_0}\alpha_{01,qq'}(t),
\label{a0n}\\
&\frac{d\alpha_{11,qq}(t)}{dt}=\frac{2}{\hbar}g(t)\,\Imm\left\{\alpha_{01,qq}(t)\;\ee^{\ii\w0 t}\right\}
+\frac{2}{\hbar}\sum_{q'}\,\Imm\left\{g_{qq'}\,\alpha_{01,q'q}(t)\;\ee^{\ii(\w0+\varepsilon_{qq'})t}\right\}
-\frac{1}{\tau_0}\,\alpha_{11,qq}(t),
\label{ann}\\
&\frac{d\alpha_{00,qq}(t)}{dt}=-\frac{2}{\hbar}g(t)\,\Imm\left\{\alpha_{01,qq}(t)\;\ee^{\ii\w0 t}\right\}
-\frac{2}{\hbar}\sum_{q'}\,\Imm\left\{g_{q'q}\,\alpha_{01,qq'}(t)\;\ee^{\ii(\w0+\varepsilon_{q'q})t}\right\}
+\frac{1}{\tau_0}\,\alpha_{11,qq}(t),
\label{a00}
\end{align}
\end{subequations}
where we have used the Hermiticity of $\hat\rho$ and $\hat{\mathcal{H}}$. The integral of eq\ \ref{a0n} can be readily written as
\begin{align}
&\alpha_{01,qq'}(t)=\frac{\ii}{\hbar}\,\alpha_{q}^0\,\alpha_{q'}^{0*}\,\int_{-\infty}^t dt'\,g(t')\,\ee^{-\ii\w0 t'-(t-t')/2\tau_0}
-\frac{1}{\hbar}\sum_{q''}g_{q''q'}\,\alpha_{q}^0\,\alpha_{q''}^{0*}\;\frac{\ee^{-\ii(\w0+\varepsilon_{q'q''})t}}{\w0+\varepsilon_{q'q''}+\ii/2\tau_0}.
\nonumber
\end{align}
At this point, we express the coupling coefficients in terms of the scatterer mode dipole $\pb_0$ through eqs\ \ref{gt} and \ref{gqq}, use the nonrecoil approximation to write $\varepsilon_{q'q''}\approx(q'-q'')v$, and convert the $q''$ sum into an integral by means of the prescription noted above. Following this procedure, we find
\begin{align}
&\alpha_{01,qq'}(t)=\int_{-\infty}^\infty\frac{d\omega}{2\pi}\,\ee^{-\ii\omega t}\,\tilde\alpha_{01,qq'}(\omega),
\nonumber
\end{align}
where
\begin{align}
&\tilde\alpha_{01,qq'}(\omega)=\frac{1}{\hbar}\,\frac{1}{\omega+\ii/2\tau_0}\;\pb_0\cdot\left[\Eb^{\rm ext}(0,\omega-\omega_0)\,\alpha_q^0\,\alpha_{q'}^{0*}
+\gb_{(\omega-\omega_0)/v}\,\alpha_q^0\,\alpha_{q'-(\omega-\omega_0)/v}^{0*}\right]
\label{alphaw}
\end{align}
and $\Eb^{\rm ext}(\rb,\omega)=\int_{-\infty}^\infty dt\,\ee^{\ii\omega t}\,\Eb^{\rm ext}(\rb,t)$.

We are interested in the time-integrated quantity
\begin{align}
T_q=\int_{-\infty}^{\infty}dt\,\alpha_{11,qq}(t)
\nonumber
\end{align}
(see below). From eq\ \ref{ann}, we find $T_q=\int_{-\infty}^\infty dt\,\ee^{-t/\tau_0}\int_{-\infty}^t dt'\,\ee^{t'/\tau_0}\,F(t')=\tau_0\int_{-\infty}^\infty dt\,F(t)$, where $F(t)$ is given by the first two terms in the right-hand side of that equation. This leads to
\begin{align}
T_q&=\frac{2\tau_0}{\hbar}\int_{-\infty}^\infty dt\;
\Imm\bigg\{
g(t)\,\alpha_{01,qq}(t)\;\ee^{\ii\w0 t}+\sum_{q'}g_{qq'}\,\alpha_{01,q'q}(t)\;\ee^{\ii(\w0+\varepsilon_{qq'})t}
\bigg\}.
\label{Tq}
\end{align}
As a first result, eq\ \ref{Tq} can help us evaluate the change in electron kinetic energy $\Delta E_{\rm el}$, starting from the variation in the population of the sample ground state due to the interaction, $\alpha_{00,qq}(\infty)-|\alpha_q^0|^2$. Multiplying this quantity by the plane wave energy $\hbar\varepsilon_q$, summing over $q$, calculating $\alpha_{00,qq}(\infty)$ from the integral of eq\ \ref{a00}, and using eq\ \ref{Tq}, we obtain
\begin{align}
\Delta E_{\rm el}&=\sum_q \hbar\varepsilon_q\,\left[\alpha_{00,qq}(\infty)-|\alpha_q^0|^2\right]
\nonumber\\ &=2\sum_{qq'}\varepsilon_{qq'} \int_{-\infty}^\infty dt\;
\Imm\big\{g_{qq'}\,\alpha_{01,q'q}(t)\;\ee^{\ii(\w0+\varepsilon_{qq'})t}\big\}.
\nonumber
\end{align}
We now convert the sum over $q'$ into an integral, change the variable of integration to $\omega=\varepsilon_{qq'}$, adopt the nonrecoil approximation $\varepsilon_{qq'}\approx(q-q')v$, identify the time integral as the Fourier transform $\tilde\alpha_{01,q'q}(\omega_0+\omega)$ (see eq\ \ref{alphaw}), and substitute the coupling coefficients from eqs\ \ref{gt} and \ref{gqq} to find
\begin{align}
\Delta E_{\rm el}&=-\frac{1}{\pi\hbar}\int_{-\infty}^\infty \omega d\omega\;
\Imm\bigg\{
\frac{1}{\omega_0+\omega+\ii/2\tau_0}
\,\left[\left(\pb_0\cdot\Eb^{{\rm ext}}(0,\omega)\right)\,\left(\pb_0\cdot\gb_{\omega/v}^*\right)\,M_{\omega/v}
+\left|\pb_0\cdot\gb_{\omega/v}\right|^2\right]
\bigg\},
\label{DE}
\end{align}
where we have used the normalization condition $\sum_q\left|\alpha_q^0\right|^2=1$ and defined
\begin{align}
M_{\omega/v}&=\sum_q\alpha_{q-\omega/v}^0\,\alpha_{q}^{0*}
=\int_{-\infty}^\infty dz\;\ee^{\ii\omega z/v}\;|\psi_\parallel(z)|^2
\label{Mwv}
\end{align}
(eq\ 9 in the main text). In the derivation of the integral in eq\ \ref{Mwv}, we have exploited the relation between $\alpha_q^0$ and $\psi_\parallel(z)$ given in eq\ \ref{psiz}. We now consider an isotropic particle characterized by three degenerate modes of transition dipoles $p_0\xx$, $p_0\yy$, and $p_0\zz$, each of them contributing to $\Delta E_{\rm el}$ with a term given by eq\ \ref{DE}, in which $\pb_0$ is substituted by the corresponding mode dipole. The sum of these contributions yields
\begin{align}
\Delta E_{\rm el}&=-\frac{1}{\pi\hbar}\int_{-\infty}^\infty \omega d\omega\;
\Imm\bigg\{
\frac{|p_0|^2}{\omega_0+\omega+\ii/2\tau_0}
\,\left[\Eb^{{\rm ext}}(0,\omega)\cdot\gb_{\omega/v}^*\,M_{\omega/v}
+\left|\gb_{\omega/v}\right|^2\right]
\bigg\}.
\nonumber
\end{align}
Finally, separating the integral in positive and negative frequency parts, and changing $\omega\rightarrow-\omega$ in the latter, we can write $\Delta E_{\rm el}=\int_0^\infty \hbar\omega\,d\omega\;d\Gamma_{\rm el}/d\omega$, where
\begin{align}\boxed{
\frac{d\Gamma_{\rm el}}{d\omega}=
-\frac{1}{\pi\hbar}\,\Imm\bigg\{
\alpha(\omega)\,\Eb^{{\rm ext}}(0,\omega)\cdot\Eb^{{\rm el}*}(\Rb_0,\omega)\,M_{\omega/v}\bigg\}
-\frac{1}{\pi\hbar}
\,\left|\Eb^{\rm el}(\Rb_0,\omega)\right|^2\,\Imm\left\{\alpha(\omega)\right\}}
\label{GEELS}
\end{align}
acts as an electron energy-change probability, in which we identify
\begin{align}
\alpha(\omega)=\frac{|p_0|^2}{\hbar}\left(\frac{1}{\omega_0-\omega-\ii/2\tau_0}+\frac{1}{\omega_0+\omega+\ii/2\tau_0}\right)
\label{polarizability}
\end{align}
as the particle polarizability \cite{PN1966} and we have renamed $\gb_{\omega/v}=\Eb^{\rm el}(\Rb_0,\omega)$ (see eq\ 8 in the main text). The opposite of the rightmost term in eq\ \ref{GEELS} coincides with the well-known expression of the EELS probability for a dipolar particle \cite{paper149}, $(4e^2\omega^2/\pi\hbar v^4\gamma^2)\left[K_1^2(\omega R_0/v\gamma)+K_0^2(\omega R_0/v\gamma)/\gamma^2\right]\Imm\{\alpha(\omega)\}$, whereas the first term arises as a result of the particle-assisted interaction between the electron and the external light field. Although we assign the latter to an $\omega>0$ component in the integral of $\Delta E_{\rm el}$, it should actually be interpreted as the net balance between energy losses and gains of energies $\pm\hbar\omega$.

Assuming a radiative decay rate $\gamma_{\rm rad}$ of the excited particle state, the number of photons emitted into the far field accumulates over time to yield
\begin{align}
\Gamma_{\rm rad}=\gamma_{\rm rad}\sum_q T_q.
\nonumber
\end{align}
We can work out this expression from eq\ \ref{Tq} by expressing $\alpha_{01,qq'}(t)$ in terms of its Fourier transform (eq\ \ref{alphaw}), following similar steps as in the derivation of eq\ \ref{DE}, and eventually summing over three orthogonal transition dipoles to describe an isotropic particle. This results in
\begin{align}
\Gamma_{\rm rad}=\frac{\gamma_{\rm rad}}{2\pi\hbar^2}\,
\int_{-\infty}^\infty d\omega\;\frac{|p_0|^2}{(\omega_0-\omega)^2+1/4\tau_0^2}\;\bigg[
|\Eb^{\rm ext}(0,\omega)|^2+|\Eb^{\rm el}(\Rb_0,\omega)|^2
+2\,\Ree\big\{\Eb^{\rm ext}(0,\omega)\cdot\Eb^{{\rm el}*}(\Rb_0,\omega)M_{\omega/v}\big\}\bigg],
\nonumber
\end{align}
where we have performed the $q$ sum by using $\sum_q\left|\alpha_q^0\right|^2=1$ and eq\ \ref{Mwv}. Neglecting for now the interference between scattered and externally incident photons ({\it i.e.}, we ignore the change in the probability of some of the decay channels stimulated by the photon population of such channels), we have \cite{paper053} $\gamma_{\rm rad}=4|p_0|^2\omega_0^3/3\hbar c^3$, so we can write $\Gamma_{\rm rad}=\int_0^\infty d\omega\;(d\Gamma_{\rm rad}/d\omega)$, where
\begin{align}\boxed{
\frac{d\Gamma_{\rm rad}}{d\omega}\approx\frac{1}{\pi\hbar}\,
\frac{2\omega^3\left|\alpha(\omega)\right|^2}{3c^3}\;\bigg[
|\Eb^{\rm ext}(0,\omega)|^2+|\Eb^{\rm el}(\Rb_0,\omega)|^2
+2\,\Ree\big\{\Eb^{\rm ext}(0,\omega)\cdot\Eb^{{\rm el}*}(\Rb_0,\omega)M_{\omega/v}\big\}\bigg]}
\label{Gammarad}
\end{align}
represents the spectrally resolved photon emission probability, in which we have assumed $\omega_0\tau_0\gg1$, approximated $\alpha(\omega)$ by the first (resonant) term in eq\ \ref{polarizability} for $\omega>0$, and taken $\omega\approx\omega_0$ in the multiplicative factors. After minor rearrangements, eq\ \ref{Gammarad} becomes eq\ 7 in the main text.

We also note that the accumulated probability of decay from the excited state of the particle is given by $\Gamma_{\rm decay}=(1/\tau_0)\sum_q T_q$, which, following the same procedure as above, is found to lead to
\begin{align}\boxed{
\frac{d\Gamma_{\rm decay}}{d\omega}\approx\frac{1}{\pi\hbar}\,
\Imm\{\alpha(\omega)\}\;\bigg[
|\Eb^{\rm ext}(0,\omega)|^2+|\Eb^{\rm el}(\Rb_0,\omega)|^2
+2\,\Ree\big\{\Eb^{\rm ext}(0,\omega)\cdot\Eb^{{\rm el}*}(\Rb_0,\omega)M_{\omega/v}\big\}\bigg].}
\label{gammadecay}
\end{align}
Importantly, in the final energy balance of the entire electron-particle-radiation system, a term
\begin{align}[box=\widebox]
\frac{d\Gamma_{\rm forward}}{d\omega}
&=-\frac{1}{\pi\hbar}\,\Imm\big\{\alpha(\omega)\big\}\,|\Eb^{\rm ext}(0,\omega)|^2
+\frac{1}{\pi\hbar}\,\Imm\big\{\alpha^*(\omega)\,\Eb^{\rm ext}(0,\omega)\cdot\Eb^{{\rm el}*}(\Rb_0,\omega)M_{\omega/v}\big\}
\label{Gfor}\\
&=-\frac{1}{\pi\hbar}\,\Imm\big\{\alpha(\omega)\,\Eb^{{\rm ext}*}(0,\omega)\cdot
[\Eb^{\rm ext}(0,\omega)+\Eb^{\rm el}(\Rb_0,\omega)M_{\omega/v}^*]\big\} \nonumber
\end{align}
is missing in order to conserve energy for each $\omega$ component according to the condition
\begin{align}\boxed{
\frac{d\Gamma_{\rm el}}{d\omega}+\frac{d\Gamma_{\rm decay}}{d\omega}+\frac{d\Gamma_{\rm forward}}{d\omega}=0.}
\label{balance}
\end{align}
We interpret $\Gamma_{\rm forward}$ as the change in photon forward emission ({\it i.e.}, toward the direction of propagation of the incident light beam) associated with interference between emitted and externally incident photons ({\it i.e.}, the type of stimulated process that we neglected in $\gamma_{\rm rad}$ above). The first term in the right-hand side of eq\ \ref{Gfor} corresponds to the depletion of the incident light, as described by the optical theorem \cite{J99} ({\it i.e.}, $(1/\pi\hbar)\Imm\{\alpha(\omega)\}\,|\Eb^{\rm ext}(0,\omega)|^2=\sigma_{\rm ext}(\omega)I(\omega)/\hbar\omega$, where $\sigma_{\rm ext}(\omega)=(4\pi\omega/c)\Imm\{\alpha(\omega)\}$ is the extinction cross section and $I(\omega)=(c/4\pi^2)|\Eb^{\rm ext}(0,\omega)|^2$ is the light intensity per unit frequency), whereas the remaining term originates in electron-light interference.

Finally, part of the energy is absorbed by the particle due to internal inelastic transitions, so the total decay of the particle excited state can be separated as
\begin{align}
\boxed{\frac{d\Gamma_{\rm decay}}{d\omega}=\frac{d\Gamma_{\rm rad}}{d\omega}+\frac{d\Gamma_{\rm abs}}{d\omega},}
\nonumber
\end{align}
where
\begin{align}\boxed{
\frac{d\Gamma_{\rm abs}}{d\omega}\approx\!\frac{1}{\pi\hbar}
\bigg[\Imm\{\alpha(\omega)\}-\frac{2\omega^3\left|\alpha(\omega)\right|^2}{3c^3}\bigg]\bigg[
|\Eb^{\rm ext}(0,\omega)|^2+|\Eb^{\rm el}(\Rb_0,\omega)|^2
+2\,\Ree\big\{\Eb^{\rm ext}(0,\omega)\cdot\Eb^{{\rm el}*}(\Rb_0,\omega)M_{\omega/v}\big\}\bigg].}
\label{Gammaabs}
\end{align}
is the spectrally resolved absorption probability. This completes our analysis of energy pathways during the interaction of the particle with external light and an incident electron. The contributions to the energy balance in eq\ \ref{balance} are thus given by eqs\ \ref{GEELS}, \ref{gammadecay}, and \ref{Gfor}, while the decay in \ref{gammadecay} can in turn be expressed as the sum of two terms corresponding to radiative and absorptive channels, as given by eqs\ \ref{Gammarad} and \ref{Gammaabs}, respectively.

We remark that the boxed equations derived above apply to isotropic dipolar particles. Repeating the same analysis without summing over three orthogonal transition dipole orientations, we obtain similar expressions for a particle characterized by a polarizability tensor $\alpha(\omega)\,\uu\otimes\uu$ ({\it i.e.}, with a single transition dipole $\pb_0$ along a direction $\uu$), for which the partial probabilities are given by eqs\ \ref{GEELS}, \ref{Gammarad}, \ref{gammadecay}, \ref{Gfor}, and \ref{Gammaabs} after substituting $\Eb^{\rm ext}$ and $\Eb^{\rm el}$ by $\uu\cdot\Eb^{\rm ext}$ and $\uu\cdot\Eb^{\rm el}$, respectively.

\subsection*{Energy Pathways from the Quantum-Electrodynamics Formalism}

The above results can be corroborated using the quantum-electrodynamics formalism developed in the Methods section of the main text. In particular, an extension of eq\ \ref{Gfor} is readily obtained by evaluating the Poynting vector along the forward direction with respect to the incident laser, assuming illumination with a well-defined incident wave vector $\kb_{\rm inc}$. Using the notation $2\Ree\left\{\Eb^{\rm ext}(0,\omega)\ee^{\ii\kb_{\rm inc}\cdot\rb-\ii\omega t}\right\}$ for the time-dependent external light electric field, the frequency-space light electric far-field ($kr\gg1$) takes the form $\Eb^{\rm light}(\rb,\omega)\approx\Eb^{\rm ext}(0,\omega)\,\ee^{\ii\kb_{\rm inc}\cdot\rb}+\fb_{\rr}^{\rm scat}(\omega)\,\ee^{\ii k r}/r$, where $k=|\kb_{\rm inc}|=\omega/c$. When inserting this expression in eq\ 33 of the main text, we can separate $d\Gamma_{\rm ff}/\Omega_{\rr}d\omega=(d\Gamma_{\rm rad}/\Omega_{\rr}d\omega)+(d\Gamma_{\rm forward}/\Omega_{\rr}d\omega)$ into the contributions coming from the $1/r$ part ({\it i.e.}, $d\Gamma_{\rm rad}/\Omega_{\rr}d\omega$, which is extensively discussed in the main text) and the remaining interference between $\Eb^{\rm ext}(0,\omega)\ee^{\ii\kb_{\rm inc}\cdot\rb}$ and $\fb_{\rr}^{\rm CL/scat}$ terms (see also eqs\ 4 in the main text). The latter generates $d\Gamma_{\rm forward}/\Omega_{\rr}d\omega$, which can be integrated over angles $\Omega_{\rr}$ following a similar asymptotic analysis as used in the derivation of the optical theorem \cite{J99}, based on the integral $\int d^2\Omega_{\rr}\,\ee^{\ii (k+\ii0^+)r-\ii\kb_{\rm inc}\cdot\rb}=2\pi\ii/kr$ (valid in the $kr\rightarrow\infty$ limit), where $k$ is supplemented by an infinitesimal imaginary part $\ii0^+$, in accordance with the retarded formalism here adopted. This leads to
\begin{align}
\frac{d\Gamma_{\rm forward}}{d\omega}=-\frac{1}{\pi \hbar k^2}{\rm Im}\left\{\Eb^{\rm ext *}(0,\omega)\cdot \left[ \fb_{\hat{\kb}_{\rm inc}}^{\rm scat }(\omega)+M^*_{\omega/v} \fb^{\rm CL}_{\hat{\kb}_{\rm inc}}(\omega)\right]\right\}, \nonumber
\end{align}
which, using eqs\ 37 and 38 in the main text, reduces to eq\ \ref{Gfor} for a dipolar particle.

Likewise, we can obtain eq\ \ref{GEELS} starting from the electron mean energy after interaction at $t\rightarrow\infty$:
\begin{align}
\Delta E_{\rm el}=\langle\hat{\mathcal{S}}^\dagger(\infty,-\infty)\hat{\mathcal{H}}_{\rm el}\hat{\mathcal{S}}(\infty,-\infty)\rangle-\langle\hat{\mathcal{H}}_{\rm el}\rangle,\nonumber 
\end{align}
where the average $\langle \cdot \rangle$ is defined as in the main text. Noticing that the interaction Hamiltonian (eq\ 25) is linear in the total current $\hat{\jb}$, we use the evolution operator (eq\ 26) and retain terms just up to quadratic order in $\hat{\jb}$ to find
\begin{align}
\Delta E_{\rm el}\approx\frac{1}{\hbar^2}\int_{-\infty}^\infty dt \int_{-\infty}^\infty dt' \left\langle\left[\hat{\mathcal{H}}_{\rm int}(t)\hat{\mathcal{H}}_{\rm el}\hat{\mathcal{H}}_{\rm int}(t')-\frac{1}{2}\left\{\hat{\mathcal{H}}_{\rm int}(t)\hat{\mathcal{H}}_{\rm int}(t'), \hat{\mathcal{H}}_{\rm el} \right\}\right]\right\rangle-\ii \left\langle \left[\hat{\chi}(\infty,-\infty),\hat{\mathcal{H}}_{\rm el}\right] \right\rangle. \label{de2order}
\end{align}
Following the same approach as in the main text, we consider the total current to be the sum of the classical laser source $\jb^{\rm ext}$ and the electron current operator $\hat{\jb}^{\rm el}$ (eq\ 29). An important technical point refers to the operator $\hat{\chi}(\infty,-\infty)=(\ii/2\hbar^2)\int_{-\infty}^\infty dt' \int_{-\infty}^{t'} dt'' \left[\hat{\mathcal{H}}_{\rm int}(t),\hat{\mathcal{H}}_{\rm int}(t')\right]$, in which only the terms that are linear in $\hat{\jb}^{\rm el}$ are not commuting with $\hat{\mathcal{H}}_{\rm el}$. In the absence of external illumination, such linear terms disappear and the remaining part of $\hat{\chi}$ gives rise to an image-potential interaction with the sample, which produces elastic diffraction of the electron, but does not change its energy \cite{paper357}. However, in the present scenario of combined electron and light interactions with the sample, $\hat{\chi}$ gives rise to changes in the electron energy, so it needs to be retained in the calculation. We now use eqs\ 18 and 20-22, together with the Onsager reciprocity relation $G_{i,i'}(\rb,\rb',\omega)=G_{i',i}(\rb',\rb,\omega)$, to rewrite eq\ \ref{de2order} as
\begin{align}
\Delta E_{\rm el}\approx \frac{4 \ii}{\hbar}\sum_{i,i'}&\int_0^\infty d\omega \int d^3 \rb \int d^3\rb' \bigg\{\ii \, {\rm Im}\big\{G_{i,i'}(\rb,\rb',\omega)\big\}\left\langle \hat{j}^{\dagger}_i(\rb,\omega)\hat{\mathcal{H}}_{\rm el}\hat{j}_{i'}(\rb',\omega) -\frac{1}{2}\left\{\hat{j}_{i}(\rb,\omega)\hat{j}_{i'}(\rb',\omega),\hat{\mathcal{H}}_{\rm el}\right\}\right\rangle \nonumber \\
&+ \frac{1}{2}{\rm Re}\big\{G_{i,i'}(\rb,\rb',\omega)\big\}\left\langle \left[\hat{j}^{\rm el \dagger}_{i}(\rb,\omega),\hat{\mathcal{H}}_{\rm el}\right] j^{\rm ext}_{i'}(\rb',\omega)+j_{i}^{\rm ext *}(\rb,\omega)\left[\hat{j}^{\rm el}_{i'}(\rb',\omega),\hat{\mathcal{H}}_{\rm el}\right] \right\rangle\bigg\}.\nonumber 
\end{align}
Finally, we evaluate the averages $\langle\cdot\rangle$ using eq\ 33 and the definition of $\hat{\mathcal{H}}_{\rm el}$. After some algebra, this leads to $\Delta E_{\rm el}=\int_0^\infty d\omega\, \hbar \omega\, d\Gamma_{\rm el}/d\omega$ with
\begin{align}
\frac{d\Gamma_{\rm el}}{d\omega}\approx  -\frac{4 e}{\hbar} \int d^3 \rb  \int d^3 \rb ' {\rm Im}\left\{ \ee^{-\ii \omega z/v} M_{\omega/v}(\Rb) \, \zz \cdot G(\rb,\rb ',\omega) \cdot \jb^{\rm ext}(\rb ',\omega) \right\} - \int d^2 \Rb\, M_0(\Rb)\, \Gamma_{\rm EELS}(\Rb,\omega)\label{genel},
\end{align}
where $M_{\omega/v}(\Rb)$ is the same as in eq\ 3 and $\Gamma_{\rm EELS}(\Rb,\omega)$ is the classical EELS probability for an electron beam focused at $\Rb$ \cite{paper149}. Equation\ \ref{genel} represents a generalization of eq\ \ref{GEELS} to arbitrary samples and incident electron wave functions. Indeed, this result reduces to eq\ \ref{GEELS} if the electron wave function can be factorized as $\psi^0(\rb)=\psi_\perp(\Rb) \psi_\parallel(z)$ with $|\psi_\perp (\Rb)|^2\approx \delta(\Rb-\Rb_0)$ ({\it i.e.}, the tightly focused beam limit) and the sample can be described by a dipolar polarizability $\alpha(\omega)$. Under such conditions, taking the particle at the origin, we can write the scattered part of the Green tensor as $G^{\rm scat}(\rb,\rb',\omega)=-4\pi\omega^2 \alpha(\omega) G^{\rm free}(\rb,\omega)\cdot G^{\rm free}(\rb',\omega)$, in terms of the free-space component $G^{\rm free}(\rb,\omega)=(-1/4\pi\omega^2)(k^2+\nabla\otimes\nabla)\left(\ee^{\ii kr}/r\right)$, and then, combining all of these elements, using the integral $\int_{-\infty}^\infty dz\, \ee^{\ii\omega(z/v+r/c)}/r=2K_0\left(\omega R/v\gamma\right)$ (see eq\ 3.914-4 in ref\ \citenum{GR1980}), and identifying $\Eb^{\rm ext}(0,\omega)=-4\pi \ii \omega \int d^3\rb\, G^{\rm free}(\rb,\omega)\cdot\jb^{\rm ext}(\rb,\omega)$, we obtain eq\ \ref{GEELS}.

\clearpage 
\pagebreak \onecolumngrid \section*{SUPPLEMENTARY FIGURES} 

\begin{figure}[H]
\centering{\includegraphics[width=0.55\textwidth]{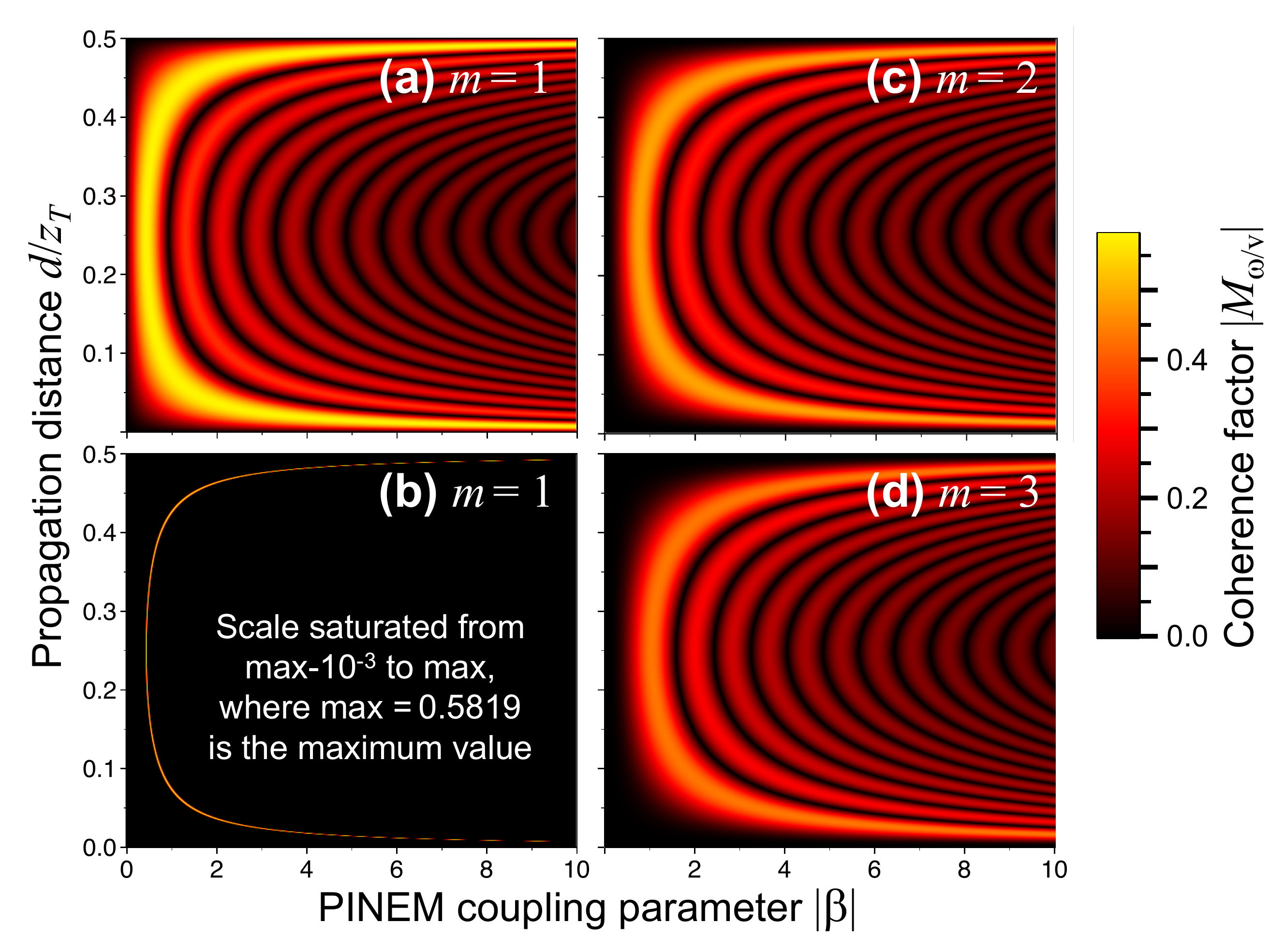}}
\caption{{\bf Coherence factor at harmonic frequencies of the PINEM laser frequency $\omega_P$.} We show plots similar to that of Figure\ 4a in the main text, but for excitation frequencies $\omega=m\omega_P$ at different harmonics $m$ of the PINEM laser frequency. Panel (a) is reproduced from Figure\ 4a in the main text for comparison. Panel (b) is extracted from panel (a) by plotting only the region corresponding to $|M_{\omega/v}|>0.580865$. The maximum values of $|M_{\omega/v}|$ are 0.581865, 0.486499, and 0.434394 for $m=1$, 2, and 3, respectively.}
\label{FigS2}
\end{figure}

\end{document}